\documentclass[longauth]{aa} 
\usepackage{morefloats}
\usepackage{graphicx}
\usepackage{xcolor,color}
\usepackage{ragged2e}
\usepackage{txfonts}
\usepackage[colorlinks=true, linkcolor=blue, citecolor=blue, urlcolor=blue]{hyperref}
\usepackage{float}

\begin{document} 

\titlerunning{Jet-driven shocks and turbulence in radio-loud AGNs}
\authorrunning{R. A. Riffel et al.}
\title{Jet-driven shocks and turbulence in radio-loud active galactic nuclei observed with JWST MIRI/MRS}

\author{Rogemar A. Riffel\inst{1,2}
\and Gabriel L. Souza-Oliveira\inst{2,1}
\and Luis Colina\inst{1}
\and Almudena Alonso-Herrero\inst{3}
\and Marina Bianchin\inst{4,5}
\and Kalliopi M. Dasyra\inst{6}
\and Lorenzo Evangelista\inst{7}
\and Kameron Goold\inst{8}
\and Pierre Guillard\inst{7}
\and Rogério Riffel\inst{9}
\and Anil Seth\inst{8} 
\and Thaisa Storchi-Bergmann\inst{9}
\and Nadia Zakamska\inst{10}
\and Samile Araujo-Santos\inst{2}
\and Anelise Audibert\inst{4,5}
\and Enrica Bellocchi\inst{11,12}
\and Steph Campbell\inst{13}
\and Françoise Combes\inst{14,15}
\and José Henrique Costa-Souza\inst{2}
\and Guilherme S. Couto\inst{16}
\and Richard I. Davies\inst{17}
\and Maitê S. Z. de Mellos\inst{2}
\and Tanio Díaz-Santos\inst{18,19}
\and Fergus R. Donnan\inst{20}
\and Ismael Garc\'{\i}a-Bernete\inst{3}
\and Santiago Garc\'{\i}a-Burillo\inst{21}
\and Laura Hermosa Muñoz\inst{3}
\and Erin K. S. Hicks\inst{22,23,24}
\and Alvaro Labiano\inst{25}
\and Enrique Lopez-Rodriguez\inst{26}
\and Vincenzo Mainieri\inst{27}
\and Christopher Packham\inst{24}
\and Miguel Pereira Santaella\inst{28}
\and Cristina Ramos Almeida\inst{4,5}
\and Lucas Ramos Vieira\inst{2,16,29}
\and Claudio Ricci\inst{30,31}
\and Vivian U\inst{32,33}
}
\institute{Centro de Astrobiología (CAB), CSIC-INTA, Ctra. de Ajalvir km 4, Torrejón de Ardoz, E-28850, Madrid, Spain
 \and Departamento de F\'isica, CCNE, Universidade Federal de Santa Maria, Av. Roraima 1000, 97105-900,  Santa Maria, RS, Brazil
 \and Centro de Astrobiología (CAB), CSIC-INTA, Camino Bajo del Castillo s/n, E-28692, Villanueva de la Cañada, Madrid, Spain  
\and Instituto de Astrof\' isica de Canarias, Calle V\'ia L\'actea, s/n, E-38205, La Laguna, Tenerife, Spain 
\and Departamento de Astrof\'isica, Universidad de La Laguna, E-38206, La Laguna, Tenerife, Spain 
\and Section of Astrophysics, Astronomy, and Mechanics, Department of Physics, National and Kapodistrian University of Athens, Panepistimioupolis Zografou, 15784
Athens, Greece 
\and Sorbonne Université, CNRS, UMR 7095, Institut d’Astrophysique de Paris, 98bis bd Arago, 75014 Paris, France 
\and Department of Physics \& Astronomy, University of Utah, Salt Lake City, UT 84112, USA 
 \and Departamento de Astronomia, Instituto de F\'\i sica, Universidade Federal do Rio Grande do Sul, CP 15051, 91501-970, Porto Alegre, RS, Brazil 
 \and Department of Physics \& Astronomy, Johns Hopkins University, Bloomberg Ctr, 3400 N. Charles St, Baltimore, MD 21218, USA 
 \and Departmento de Física de la Tierra y Astrofísica, Fac. de CC Físicas, Universidad Complutense de Madrid, E-28040 Madrid, Spain 
\and Instituto de Física de Partículas y del Cosmos IPARCOS, Fac. CC Físicas, Universidad Complutense de Madrid, E-28040 Madrid, Spain  
\and School of Mathematics, Statistics, and Physics, Newcastle University, Newcastle upon Tyne NE1 7RU, UK 
\and Observatoire de Paris, LUX, PSL University, Sorbonne Université, CNRS, F-75014 Paris, France 
\and Collège de France, 11 Place Marcelin Berthelot, 75231 Paris, France 
\and Leibniz-Institut für Astrophysik Potsdam, An der Sternwarte 16, 14482 Potsdam, Germany 
\and Max-Planck-Institut für Extraterrestrische Physik, Postfach 1312,
85741 Garching, Germany 
\and Institute of Astrophysics, Foundation for Research and TechnologyHellas, 71110 Heraklion, Greece 
\and School of Sciences, European University Cyprus, Diogenes Street, Engomi 1516, Nicosia, Cyprus 
\and Department of Astrophysics, University of California San Diego, 9500 Gilman Drive, San Diego, CA 92093, USA 
\and Observatorio de Madrid, OAN-IGN, Alfonso XII, 3, E-28014 Madrid, Spain  
\and Department of Physics \& Astronomy, University of Alaska Anchorage, Anchorage, AK 99508-4664, USA 
\and Department of Physics, University of Alaska, Fairbanks, Alaska 99775-5920, USA 
\and Department of Physics \& Astronomy, The University of Texas at San Antonio, One UTSA Circle, San Antonio, TX 78249, USA 
\and Telespazio UK for the European Space Agency (ESA), ESAC, Camino Bajo del Castillo s/n, 28692, Villanueva de la Cañada, Spain 
\and Department of Physics \& Astronomy, University of South Carolina, Columbia, SC 29208, USA 
\and European Southern Observatory, Karl-Schwarzschild-Strasse 2, Garching bei München, Germany 
\and Instituto de Física Fundamental, CSIC, Calle Serrano 123, E-28006 Madrid, Spain 
\and Instituto Federal Catarinense - Campus Concórdia, 89703-720 Concórdia, SC, Brazil
\and Instituto de Estudios Astrofísicos, Facultad de Ingeniería y Ciencias, Universidad Diego Portales, Av. Ejército Libertador 441, Santiago, Chile 
\and Department of Astronomy, University of Geneva, ch. d’Ecogia 16, 1290, Versoix, Switzerland 
\and Department of Physics and Astronomy, 4129 Frederick Reines Hall, University of California, Irvine, CA 92697, USA 
\and IPAC, Caltech, 1200 E. California Blvd., Pasadena, CA 91125, USA
}


 
  \abstract
  {Jet--cloud interactions are a key manifestation of active galactic nucleus (AGN) feedback on nuclear scales, which is distinct from the large-scale radio-mode feedback that suppresses gas cooling in galaxy halos. On these smaller scales, radio jets can inject energy and momentum into the interstellar medium (ISM), shaping the physical and kinematic properties of the nuclear and circumnuclear regions of galaxies. 
  Using JWST MIRI/MRS observations of seven nearby radio-loud AGNs (3C\,293, 3C\,305, Centaurus\,A, Cygnus\,A, IC\,5063, NGC\,1052, and M\,87), we investigated jet-driven turbulence in both the warm molecular and ionized gas phases. By combining spatially resolved H$_2$/polycyclic aromatic hydrocarbon flux ratios with diagnostic line ratios of the ionized gas, we constrained the dominant H$_2$ excitation processes and assessed the impact of radio jet--ISM interactions on the multiphase gas.
  We find that radio jets drive increased turbulence in both molecular and ionized (traced by [Fe\,{\sc ii}], [Ne\,{\sc ii}], and [Ne\,{\sc iii}] lines) gas, not only along but also perpendicular to the jet axis, indicating that jet--ISM interactions extend beyond the collimated jet channel and affect the nuclear environment. Strong correlations between the H$_2$/polycyclic aromatic hydrocarbon ratio, the H$_2$ excitation temperature, and shock-sensitive ionized-gas tracers indicate that jet-driven shocks dominate the excitation of the H$_2$ rotational lines in most sources. These results indicate that radio jets are a key driver of multiphase ISM kinematics and excitation in nearby radio-loud galaxies.
}

   \keywords{galaxies: active -- galaxies: jets -- galaxies: ISM -- galaxies: kinematics and dynamics}

   \maketitle
%

\section{Introduction}

The interplay between active galactic nuclei (AGNs) and star formation (SF) in galaxies is a fundamental process in galaxy evolution, and potentially links the growth of supermassive black holes with their host galaxies. AGN feedback operates through multiple channels, including relativistic jets launched from the inner accretion disk, winds driven from its outer regions, and intense radiation produced by the hot gas in the disk or its corona \citep{elvis00,ciotti10,Harrison24}. In particular, radio jets inject mechanical energy into the surrounding halo, preventing gas cooling and thereby suppressing SF in the host galaxy \citep[i.e., negative feedback;][]{cavagnolo10,Mukherjee25}.  At the same time, AGN feedback can also trigger SF, producing so-called positive feedback. In particular, the interaction of jets and outflows with the interstellar medium (ISM) can enhance SF by compressing molecular clouds or by triggering SF within the outflowing material itself \citep[i.e., positive feedback;][]{Silk13,Cresci15,Maiolino17,Bessiere22,Hermosa24}.

Molecular hydrogen (H$_2$) and polycyclic aromatic hydrocarbons (PAHs) are key tracers of the colder phases of the ISM in nearby AGN hosts. While PAH emission is commonly associated with photodissociation regions and SF \citep[e.g.,][]{Tielens08,Esquej14,Hrodmarsson25}, H$_2$ rotational and ro-vibrational lines probe the warm and hot molecular gas and are sensitive to a variety of excitation mechanisms, including UV radiation \citep[e.g.,][]{Sternberg89,davies03}, X-rays \citep[e.g.,][]{maloney96}, shocks \citep[e.g.,][]{Hollenbach89,hill14,kristensen23}, and nonthermal UV fluorescence \citep[e.g.,][]{black87}. In AGN environments, shock excitation related to radio jets and outflows can play a dominant role in powering the H$_2$ emission, particularly in regions where PAH emission is weak or suppressed \citep{Ogle07,Ogle10,Guillard09,Guillard12}.

Observations with the \textit{James Webb} Space Telescope (JWST) Mid-Infrared Instrument (MIRI) Medium Resolution Spectrometer (MRS) have significantly improved our ability to probe spatial variations in the H$_2$/PAH flux ratio, providing new insights into the origin of the molecular gas emission in nearby AGN hosts. For instance, \citet{Garcia-Bernete24} present spatially resolved H$_2$/PAH maps of three nearby Seyfert galaxies (NGC~5506, NGC~5728, and NGC~7172) from the Galaxy Activity, Torus, and Outflow Survey \citep[GATOS; e.g.,][]{Garcia-Burillo21,Alonso-Herrero21,Davies24,Garcia-Bernete24b}. They find that the H$_2$/PAH ratio is enhanced within the AGN ionization cones, and concluded that AGN illumination and coupling can affect the PAH population on both nuclear and kiloparsec scales.  \citet{rogemar25_jwst} present H$_2$/PAH ratio maps of three nearby AGN hosts (3C~293, CGCG~012$-$070, and NGC~3884) selected from \textit{Spitzer} observations as some of the most extreme cases of H$_2$ emission excess relative to that expected for star-forming galaxies. They conclude that shocks associated with outflows and radio jet--ISM interactions play a key role in powering the observed H$_2$ emission. 
\citet{Delaney25,Delaney26} used JWST MIRI/MRS observations of nearby Seyfert galaxies from GATOS \citep{Garcia-Burillo21,Alonso-Herrero21} to investigate the origin of the H$_2$ emission.  Their results show that the dominant excitation mechanism of the molecular gas varies among sources, with contributions from both AGN irradiation and shock heating. In addition, they report evidence of interactions between the AGN ionization cone, possible outflows, and the rotating molecular disk, indicating that outflow-driven shocks coupled with ionization cone irradiation play a significant role in heating the H$_2$. There have also been studies focusing on individual radio-loud (RL) galaxies; however, a systematic investigation of this population using MIRI/MRS is still lacking, leaving the role of radio jets in shaping the H$_2$ emission in RL AGNs poorly constrained.

In this work, we used JWST MIRI/MRS data of seven RL sources to investigate the origin of the molecular gas emission and its connection with the radio jets, by separating regions aligned with and perpendicular to the radio jet axis with the goal of identifying general trends between gas emission and the orientation of the radio jets. This paper is organized as follows: In Sect.~\ref{sec:data} we describe the data processing and measurements. In Sect.~\ref{sec:res} we present the results, focusing on the H$_2$/PAH ratio and the relation between the radio jet orientation, line intensity ratios, and gas kinematics. In Sect.~\ref{sec:disc} we discuss these results, and in Sect.~\ref{sec:conc} we summarize the main conclusions.

\section{Data and measurements} \label{sec:data}

\subsection{Sample of radio-loud AGNs and JWST MIRI/MRS data}

\begin{table*}
\centering
\caption{Basic properties of the sample.} \label{tab:sample}
\begin{tabular}{|lcccccccccl|}
\hline
(1)   & (2) & (3) & (4)                    & (5)           & (6)            & (7)  & (8) & (9)& (10) & (11)\\
Name & Morphology & $z$ & $D$ & $\log L_{\mathrm{1.4}}$ & $q_{\rm 22}$  & PA$_{\rm rad}$ & $\Psi_{\rm 0}$ & $i$ & PID & PI \\
 &  & & [Mpc] & [W Hz$^{-1}$] &  & [deg] & [deg] & [deg] & & \\
\hline
3C\:293 & S? & 0.0451 & 200.2 & 25.4$^{\rm e}$ & $-2.15$ & 100$^{\rm i}$           &45 & 58& 1924 & Riffel, R. A.\\

3C\:305 & SB0 & 0.0417 & 175.1 & 25.0$^{\rm e}$ & $-1.83$ & 54$^{\rm j}$           &70 & 48& 4237 & Ogle, P.\\
Centaurus\:A & S0 pec & 0.0018 & 3.1$^{\rm a}$ & 23.5$^{\rm f}$ & $-1.64$ &  51$^{\rm k}$ &155& 60&  1269 & Luetzgendorf, N.\\
Cygnus\:A & S? & 0.0561 & 234.3 & 28.0$^{\rm g}$ & $-3.38$ & 284$^{\rm l}$ &  30& 70& 4065 & Ogle, P. \\
IC\:5063 & SA0$^+$(s)? & 0.0118 & 37.5$^{\rm b}$ & 23.5$^{\rm g}$ & $+0.21$& 115$^{\rm m}$ &115 & 80 & 2004  & Dasyra, K. \\
NGC\:1052 & E4 & 0.0053 & 18.0$^{\rm c}$ & 22.6$^{\rm h}$ & $-0.32$ & 85$^{\rm n}$ &5&  75& 2016 & Seth, A. \\
M\:87 & cD pec & 0.0044 & 15.8$^{\rm d}$ & 24.6$^{\rm h}$ & $-3.27$ & 290$^{\rm o}$  &153& 31& 2016 & Seth, A. \\
\hline
\end{tabular}
\tablefoot{(1) Galaxy name; (2) Morphological classification from \citet{Vaucouleurs91}; (3) Redshift; (4) Distance. For 3C\:293, 3C\:305 and Cygnus\:A, the distances are estimated from the redshift, assuming a cosmology with $h = 0.7$, $\Omega_{m} = 0.3$, and $\Omega_{\Lambda} = 0.7$. References for the other galaxies: (a): \citet{Majaess08}; (b) \cite{Theureau07}; (c) \citet{Jensen03}; (d) \citet{Oldham16}; (5) 1.4 GHz luminosity. References: (e) \citet{Condon02}; (f) \citet{Condon96}; (g) \citet{Birzan04}; (h) \citet{Allison14}; (6) The $q_{\rm 22}$ parameter, defined as 
$ q_{\rm 22} \equiv \log\left(\frac{S_{22}}{S_{1.4}}\right)$,
where $S_{22}$ (from the WISE $W4$ magnitude; \citealt{wise14}) and $S_{1.4}$ are the flux densities at 22~$\mu$m and 1.4~GHz, respectively; (7) Orientation of the large-scale radio jet. References: (i) \citet{Kukreti22}; (j) \citet{Jackson03}; (k) \citet{Clarke92}; (l) \citet{Carilli96}; (m) \citet{Morganti07}; (n) \citet{Kadler04}; (o) \citet{Biretta91}; (8) Orientation of the galaxy's disk or dust lane;  (9) inclination of the disk; References for $\Psi_0$ and $i$ are: 3C\:293 \citep{labiano14}, 3C\:305 \citep{Morganti23}, Centaurus~A \citep{Espada17}, Cygnus~A \citep{Carilli22}, IC\:5063 \citep{Morganti15}, NGC\:1052 \citep{Kameno20}, and M\:87 \citep{Boizelle25}, based on CO observations.  (10)  JWST program ID; and (11) principal investigator of the proposal. 
}
 \end{table*}

We used archival MIRI/MRS \citep{2015PASP..127..646W,Labiano21,2023A&A...675A.111A}  data of RL AGNs, retrieved from the Mikulski Archive for Space Telescopes (MAST) Portal for JWST, selecting AGNs at $z<0.1$ with full spectral coverage of the MRS spectral range (short, medium, and long sub-bands).
The redshift limit was chosen to allow the study of gas emission and kinematics on scales of hundreds of parsecs. The resulting sample comprises seven galaxies, listed in Table~\ref{tab:sample}. The RL AGNs are identified by an excess of radio emission relative to the IR–radio correlation observed in star-forming galaxies and radio-quiet AGNs. Specifically, we used the $q_{\rm 22}$ parameter, defined as  
$q_{\rm 22} \equiv \log\left(S_{22}/S_{1.4}\right)$,  
where $S_{22}$ and $S_{1.4}$ are the flux densities at 22~$\mu$m and 1.4~GHz, respectively, with RL AGNs typically showing $q_{\rm 22} < 0$ \citep{Radcliffe21}. The total 1.4~GHz flux densities are taken from the references listed in Table~\ref{tab:sample}, while $S_{22}$ is obtained from the WISE $W4$ magnitude \citep{wise14}.  In addition, we included the borderline galaxy IC\,5063 ($q_{22} = 0.21$), which is known to host jet–cloud interactions, as revealed by detailed multiwavelength studies \citep[e.g.,][]{Dasyra15,Dasyra16,Dasyra24}. Notably, a less conservative cut of $q_{22}=0.5$ is often adopted to define RL AGNs \citep[e.g.,][]{Bonzini13,Audibert22}. All seven galaxies are also classified as RL AGNs based on their excess radio emission relative to X-ray and optical fluxes, i.e.,  $
\log F_{\rm 5\,GHz}/F_B > 1$ \citep{Kellermann89} and 
$ \log F_{\rm 5\,GHz}/F_{\rm 2-10\,keV} > -4.9$ \citep{Terashima03},
where $F_{\rm 5\,GHz}$, $F_B$, and $F_{\rm 2-10\,keV}$ are the flux densities at 5~GHz, in the $B$ band, and in the 2–10~keV X-ray band, respectively. Finally, we note that NGC\,1052 hosts a low-power radio jet, the faintest among the galaxies in our sample, and exhibits little or no ongoing star formation \citep{Dahmer-Hahn19b,Rogerio22}.

The data were processed with JWST Science Calibration Pipeline version 1.19.1 \citep{bushouse_2024}, using the reference file \texttt{jwst\_1303.pmap}. The data cubes are provided by the archive as individual band cubes for each target. These band cubes are aligned and combined using their native spatial sampling into channel cubes, which are first convolved to match the JWST/MIRI point spread function at 15~$\mu$m \citep[corresponding to $\sim$0.6~arcsec;][]{Law23}, ensuring a uniform angular resolution across all cubes. All channels for each target are then aligned to a common WCS and resampled onto a uniform spatial grid with a pixel scale of 0.2~arcsec. Next, the cubes are interpolated along the spectral axis using the Channel 1 spectral resolution as reference, producing final combined and interpolated cubes with consistent spectral sampling. Finally, the fluxes are converted to units of erg\,s$^{-1}$\,cm$^{-2}$\,\AA$^{-1}$, accounting for the new nominal pixel area. This procedure results in a single datacube per galaxy.

We employed nonparametric measurements to derive the emission-line fluxes and velocity dispersions, following a similar procedure as in \citet{rogemar26_CAB}. Measurements were performed for the following emission lines: H$_2$ S(1)\,17.04$\mu$m, S(3)\,9.66$\mu$m, and S(5)\,6.91$\mu$m; [Fe\,{\sc ii}]\,5.34\,$\mu$m; [Ne\,{\sc ii}]\,12.81\,$\mu$m and [Ne\,{\sc iii}]\,15.56\,$\mu$m. The flux of each line is obtained by integrating the continuum-subtracted spectra within velocity channels ranging from 2000 to 4000~km\,s$^{-1}$, centered at the expected wavelength of each transition. The underlying continuum is modeled and subtracted by fitting a linear function to adjacent spectral windows, free of emission-line contribution.  We estimated the gas velocity dispersion using the $W_{\rm 80}$ parameter, i.e., the velocity width that encloses 80\% of the total line flux, and defined, as $W_{\rm 80} \equiv V_{\rm 90} - V_{\rm 10}$, where $V_{\rm 90}$ and $V_{\rm 10}$ correspond to the velocities at which 90\% and 10\% of the cumulative line flux are reached, respectively. The measured $W_{\rm 80}$ values were corrected for instrumental broadening  \citep{Jones23}. 
The width of the PAH\,11.3$\:\mu$m feature is due to the range of energies of the bending mode of different molecules, and it is much broader than the range of physical velocities in the gas. Therefore, the integration is carried out over wavelength intervals with widths ranging from 0.3 to 0.6 $\mu$m, centered at 11.3~$\mu$m. The equivalent width (EW) of the PAH feature is then computed as the ratio between the integrated flux and the underlying continuum level. For all measurements, we only considered spaxels where the corresponding emission feature is detected with an amplitude-to-noise ratio (A/N) greater than 5. The A/N is estimated by dividing the emission-line amplitude by the standard deviation of the continuum in adjacent spectral windows. The PAH\,11.3$\:\mu$m feature is not detected in any spaxel of M\,87.

\subsection{Complementary Spitzer  data}

For complementary analysis, we used spectroscopic data from the \textit{Spitzer} Space Telescope. Specifically, we employed the PAH feature fluxes and EWs, as well as emission-line fluxes, provided by \citet{Spoon22}, who compiled the spectra of 3335 galaxies observed in the low-resolution staring mode of the Infrared Spectrograph (IRS), covering the spectral range from 5.4 to 36~$\mu$m. From this dataset, we defined AGN and star-forming subsamples, which we compared with the spatially resolved JWST results for the galaxies studied here. Due to the wide range in the distances of the galaxies, the fixed size of the \textit{Spitzer} slit samples different physical sizes and therefore the derived luminosity does not correspond to total, integrated luminosities. However, emission-line and PAH ratios within the slit remain reliable if the emitting regions have similar extent. Therefore, we used these data to compare line ratios, rather than luminosities, with the JWST measurements.

\section{Results}\label{sec:res}

Detailed studies of individual sources based on MIRI/MRS have already been conducted for most objects in our sample \citep{henrique24,Dasyra24,Goold24,Alonso-Herrero25,Ogle25,rogemar25_jwst,Evangelista26}. Here, we focused our analysis on the \(\mathrm{H_2/PAH}\) ratio and on the relation between the jet orientation and the gas kinematics.

In Appendix~\ref{appendix:2D} we present the flux and $W_{\rm 80}$ maps of the H$_2$ S(3) line tracing warm molecular gas \citep{Rigopoulou02,Roussel07}, [Fe\,{\sc ii}] 5.34$\:\mu$m frequently associated with shocks in partially ionized regions \citep{Forbes93,Hartigan04}, and [Ne\,{\sc iii}] 15.56$\:\mu$m tracing the AGN radiation field \citep{Melendez08,Weaver10,Pereira-Santaella10}. Although [Fe\,{\sc ii}] can also be produced by photoionization from the central AGN in some cases \citep[e.g.,][]{Dors12,Ogle25}, and [Ne\,{\sc iii}] can include a contribution from shocks \citep[e.g.,][]{RamosAlmeida25}, the combination of these tracers provides a consistent view of the dominant excitation mechanisms. In addition, we present the PAH\,11.3$\:\mu$m flux and EW maps, H$_2$\,S(3)/PAH\,11.3$\:\mu$m and \,[Ne\,{\sc iii}]\,15.56$\:\mu$m/[Ne\,{\sc ii}]\,12.81$\:\mu$m line ratio maps, and  H$_2$ temperature maps using the flux ratios of the S(3) and S(1) lines [$T_{\rm H2}{\rm (3,1)}$], which trace the lower-excitation warm molecular gas, and the S(3) and S(5) lines [$T_{\rm H2}{\rm (5,3)}$], which probe the higher-excitation component. The excitation temperatures were derived assuming local thermodynamic equilibrium, using Boltzmann excitation diagrams \citep[e.g.,][]{Goldsmith99,Roussel07}. These diagnostics are used to investigate spatial correlations between the \(\mathrm{H_2/PAH}\) ratio, the radio jet, AGN ionization, and shocked gas emission. Besides a detailed description of these maps, Appendix~\ref{appendix:2D} also presents a brief contextualization within the framework of previous radio and JWST-based results for the galaxies in our sample.  In the following, we describe general trends between emission-line ratios and gas kinematics as a function of the orientation of the radio jet.

\subsection{Relation between H$_2$/PAH and the PAH equivalent width} \label{sec:ew}

   \begin{figure}
   \centering
   \includegraphics[width=0.49\textwidth]{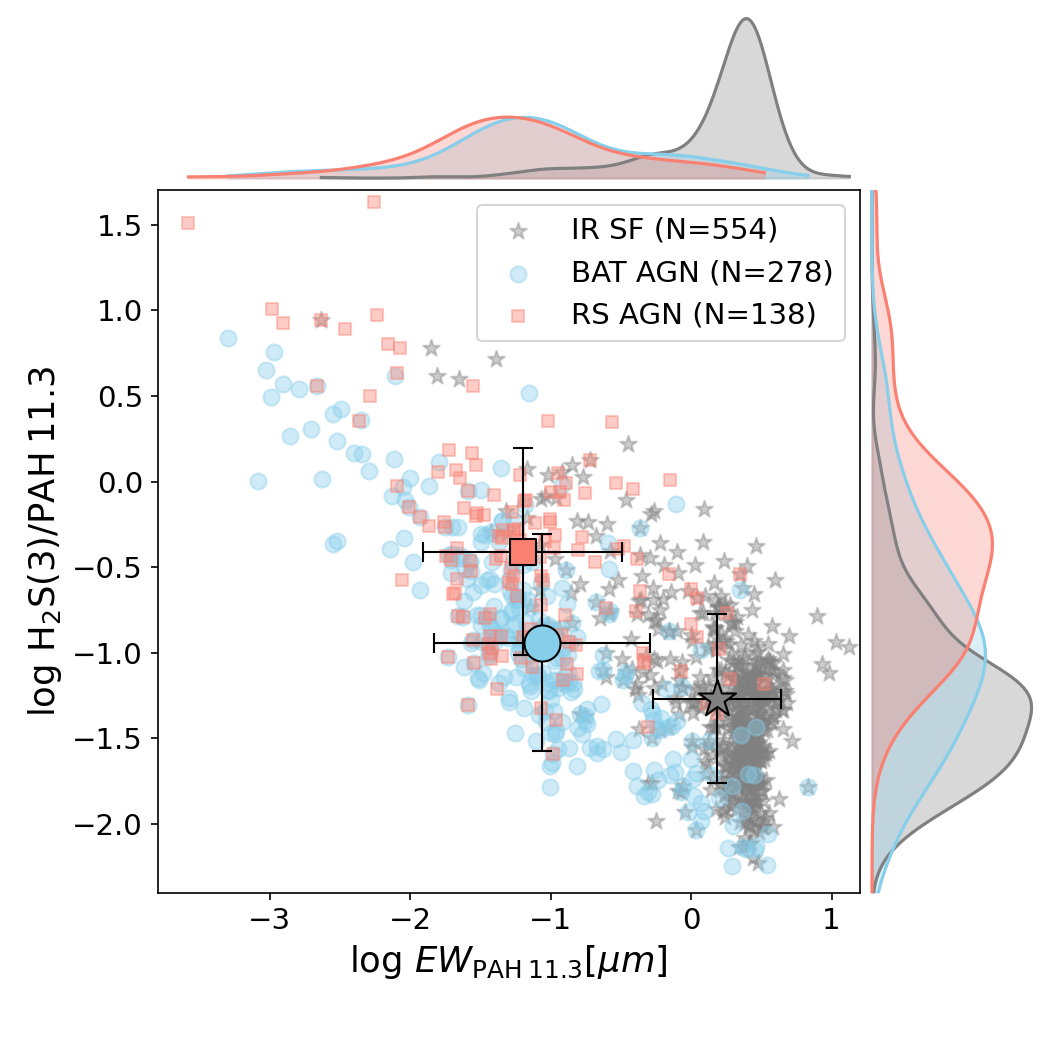} 
    \includegraphics[width=0.49\textwidth]{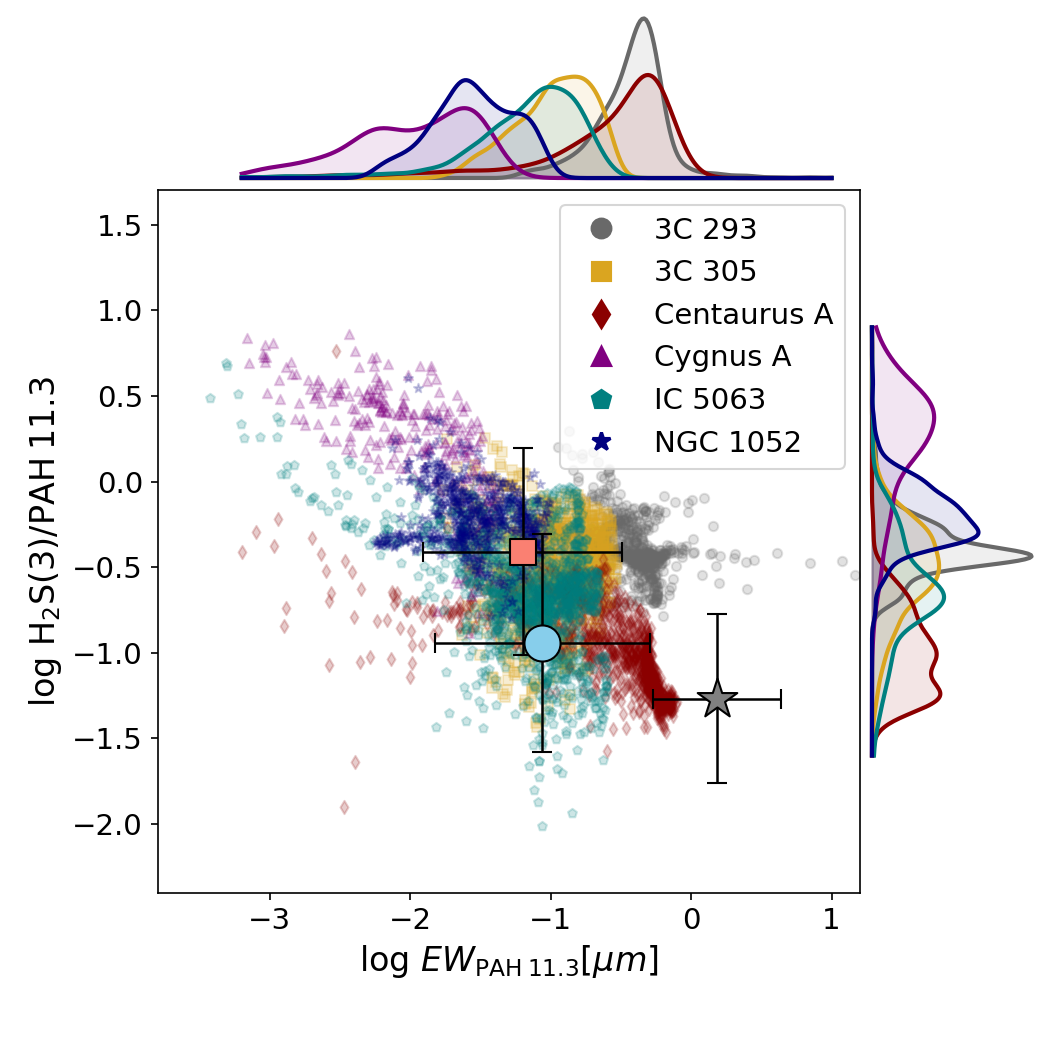} 
      \caption{Plots of $\log\,$H$_2$\,S(3)/PAH\,11.3 versus $\log\,$$EW$$_{\rm PAH\,11.3}$ using \textit{Spitzer} (\textit{top}) and MIRI/MRS data (\textit{bottom}). In the top panel, the small points in different colors represent individual galaxies from the SF, BAT AGN, and RS AGN subsamples. The bottom panel presents the JWST MRS spaxel-by-spaxel measurements for each galaxy in our sample. In both panels, the larger markers with error bars indicate the mean values and standard deviations for each \textit{Spitzer} subsample. The histograms on the top and right display the distributions of $\log\,$$EW$$_{\rm PAH\,11.3}$ and $\log\,$H$_2$\,S(3)/PAH\,11.3, respectively, following the same color scheme used for the points.}
         \label{fig:H2PAH_vs_EWPAH}
   \end{figure}

To ensure a consistent comparison between our spatially resolved (spaxel-by-spaxel) measurements and the integrated properties derived for a larger galaxy sample, we cross-matched the \textit{Spitzer} dataset from \citet{Spoon22} with the 105-month \textit{Swift} Burst Alert Telescope (BAT) survey catalog of hard X-ray (14--195~keV) sources \citep{BAT105}, with the Very Large Array (VLA) Faint Images of the Radio Sky at Twenty-Centimeters (FIRST) survey \citep{Becker1995,Lofthouse18,Lin18}; and with the Wide-field Infrared Survey Explorer \citep[WISE;][]{Wright10}.  This approach is employed to define samples of X-ray-selected AGNs (BAT AGNs) and radio-selected (RS) AGNs (AGNs), drawn from the BAT and FIRST surveys, respectively, as well as a comparison sample of star-forming galaxies (IR SF) identified using WISE observations.  The hard X-ray emission is among the most reliable tracers of AGN activity, as it predominantly probes the intrinsic radiation from the central engine \citep{Tueller08,Ricci17}. Moreover, it is far less affected by line-of-sight obscuration than optical or soft X-ray wavelengths \citep{Ricci15}.  The RS AGN sample is defined based on the 1.4 GHz power using the threshold $P_{\rm 1.4 GHz}=10^{24}\:{\rm W\:Hz^{-1}}$ \citep[e.g.,][]{Tadhunter16}. The star-forming galaxy sample is defined based on the WISE $W1-W2$ and $W2-W3$ colors with the following cuts: $W1-W2<0.5$ and $2.0<W2-W3<4.0$, values typical of star-forming spiral galaxies \citep{Wright10,Stern12}.

 The EWs of PAH features are widely used to distinguish AGNs from star-forming galaxies, with AGNs generally showing values lower than star-forming galaxies \citep[e.g.,][]{Spoon07,Diamond-Stanic10,Alonso-Herrero14}, except for extremely obscured AGNs \citep[e.g.,][]{garcia-bernete22}. In addition, the H$_{2}$/PAH ratio in AGN hosts is systematically higher than that measured in purely star-forming galaxies, and the most extreme values are found among RS AGNs \citep{Roussel07,Ogle10,lambrides19,Hermosa26}. Figure~\ref{fig:H2PAH_vs_EWPAH} shows the plots of $\log\,$H$_2$\,S(3)/PAH \,11.3 versus $\log\,$$EW$$_{\rm PAH\,11.3}$ based on the \textit{Spitzer} data (top panel) and on the JWST MIRI/MRS data (bottom panel). The \textit{Spitzer}-based results show $EW$$_{\rm PAH\,11.3}$ systematically larger for the IR SF sample than for the AGN samples, as expected, but the AGN samples show much broader distribution of values, with radio- and X-ray- selected AGNs presenting very similar mean values. The H$_2$/PAH ratio exhibits distinct distributions among the different galaxy samples, with the lowest values found in star-forming galaxies, intermediate values in X-ray selected AGNs, and the highest ratios observed in RS AGNs.   There is an overlap between AGNs, particularly X-ray selected sources, and star-forming galaxies at the high end of the $EW$$_{\rm PAH\,11.3}$ and ${\rm H_2}$/PAH distributions, which is likely associated with the large aperture of the \textit{Spitzer} data, encompassing emission from circumnuclear star-forming regions commonly found around AGNs \citep[e.g.,][]{Colina02,Munoz07,Sales10,Ruschel-Dutra17,rogemar_N4303,Hennig18}. 

The bottom panel of Fig.~\ref{fig:H2PAH_vs_EWPAH} shows the $\log\,$H$_2$\,S(3)/PAH\,11.3 versus $\log\,$$EW$$_{\rm PAH\,11.3}$  plot for the galaxies in our sample using spaxel-by-spaxel measurements. Each galaxy is represented by different symbols and colors as indicated in the figure label. Consistent with what is observed using integrated spectra, there is a strong anticorrelation between $EW$$_{\rm PAH\,11.3}$  and ${\rm H_2}$/PAH. Most points lie within the region occupied by AGN hosts in the \textit{Spitzer}-based diagram, suggesting that spaxel-by-spaxel measurements of these parameters may serve as reliable AGN selection criteria, similarly to integrated measurements.

   \begin{figure*}
   \centering
\includegraphics[width=0.3\textwidth]{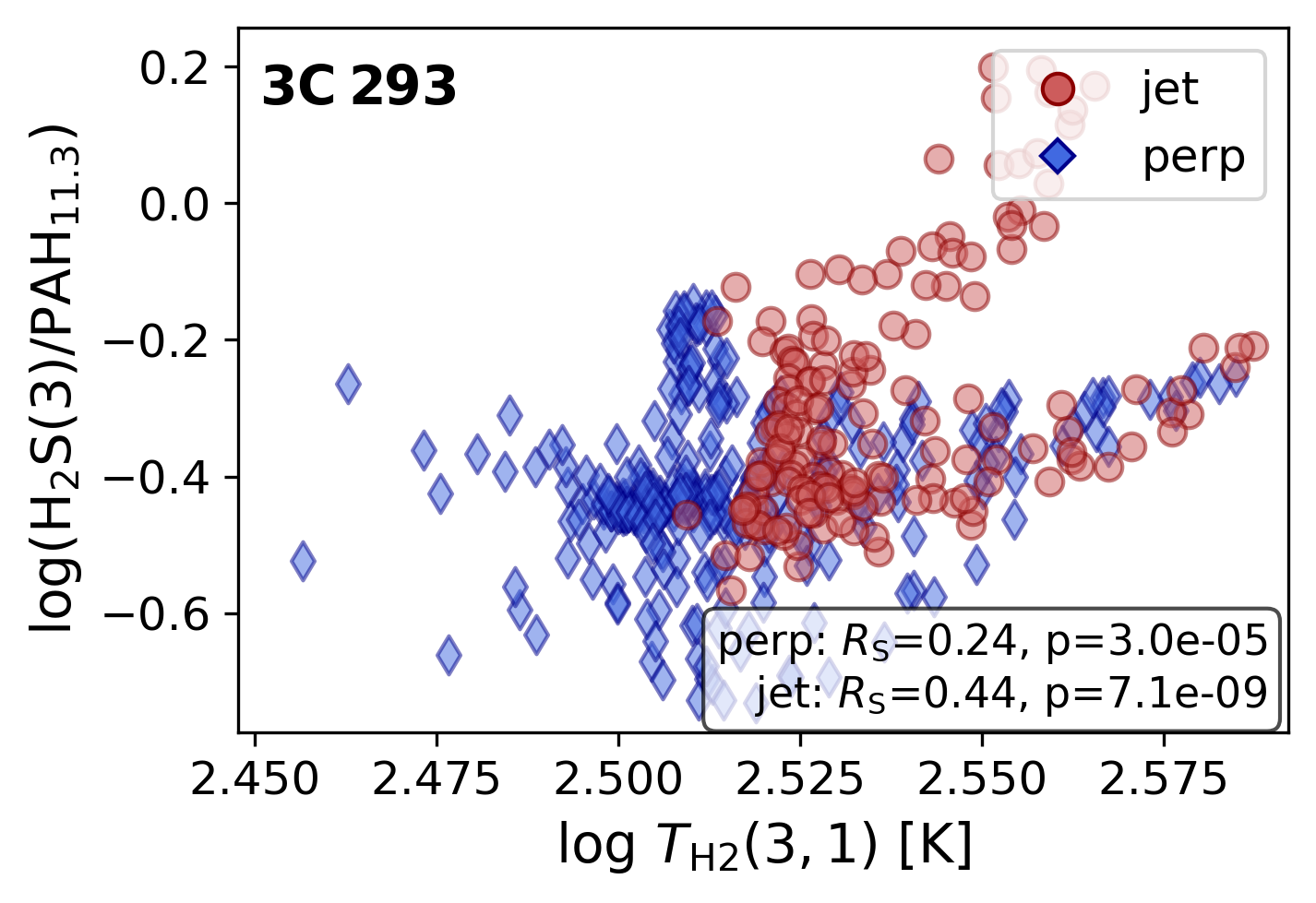} 
\includegraphics[width=0.3\textwidth]{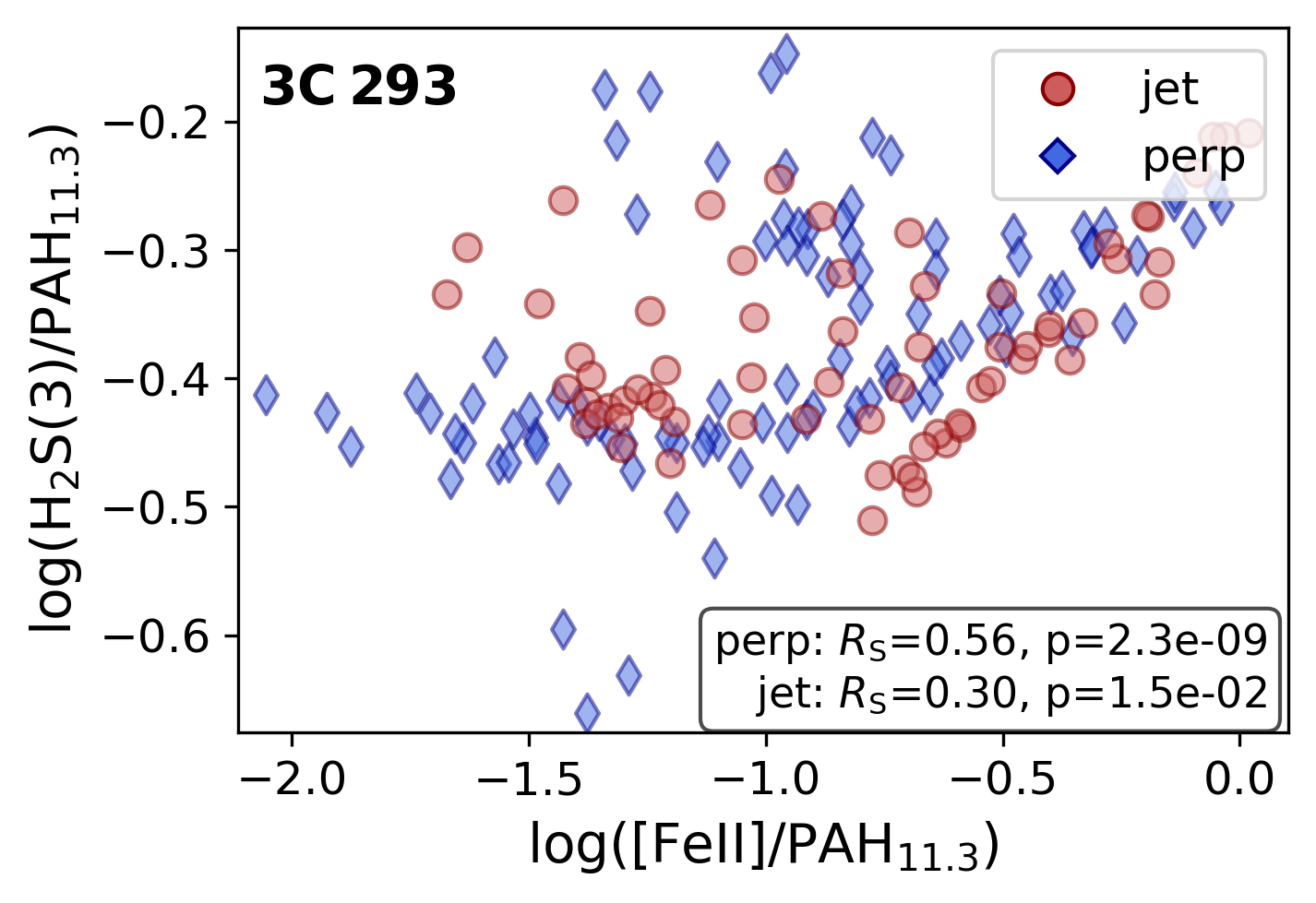} 
\includegraphics[width=0.3\textwidth]{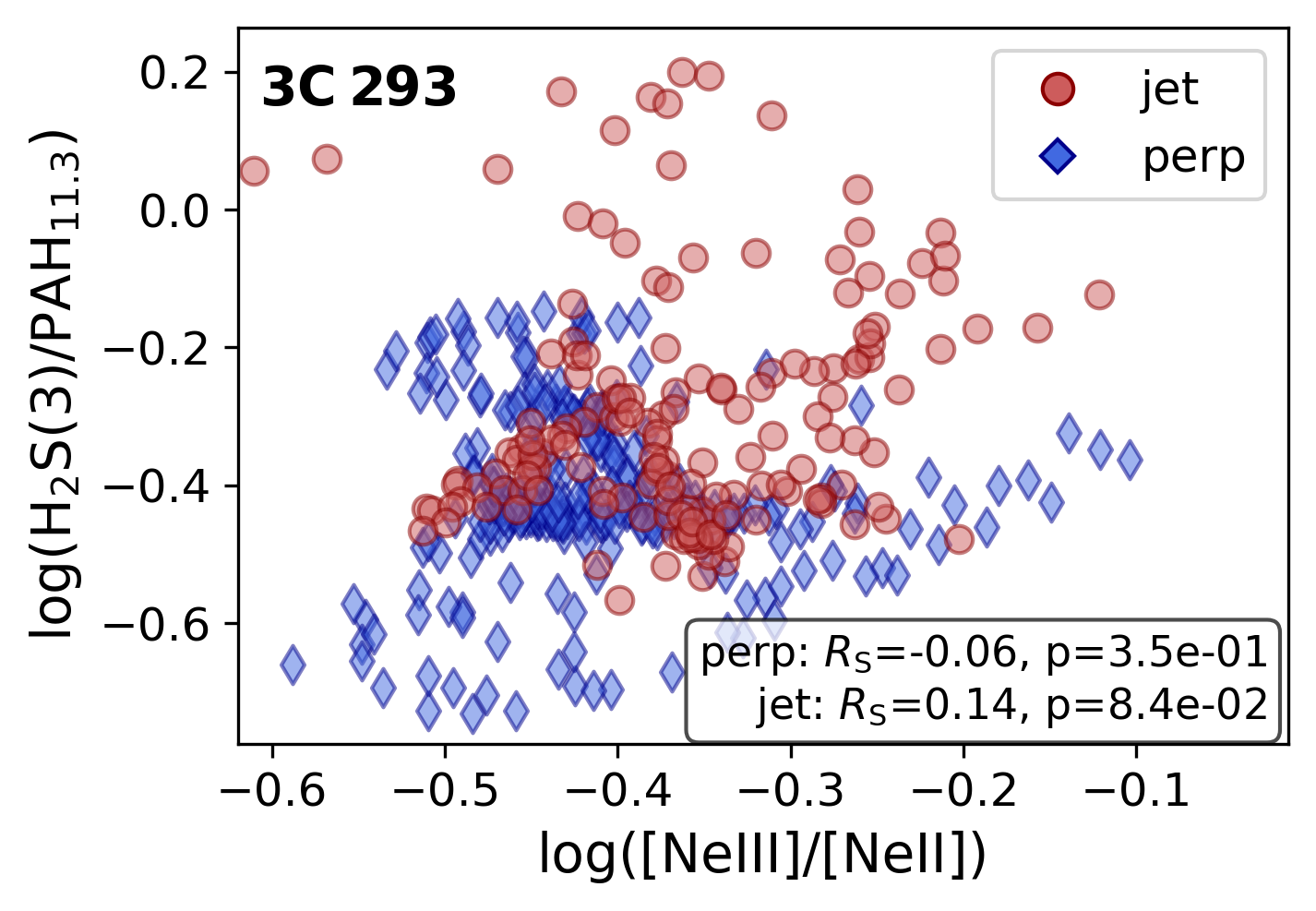}

\includegraphics[width=0.3\textwidth]{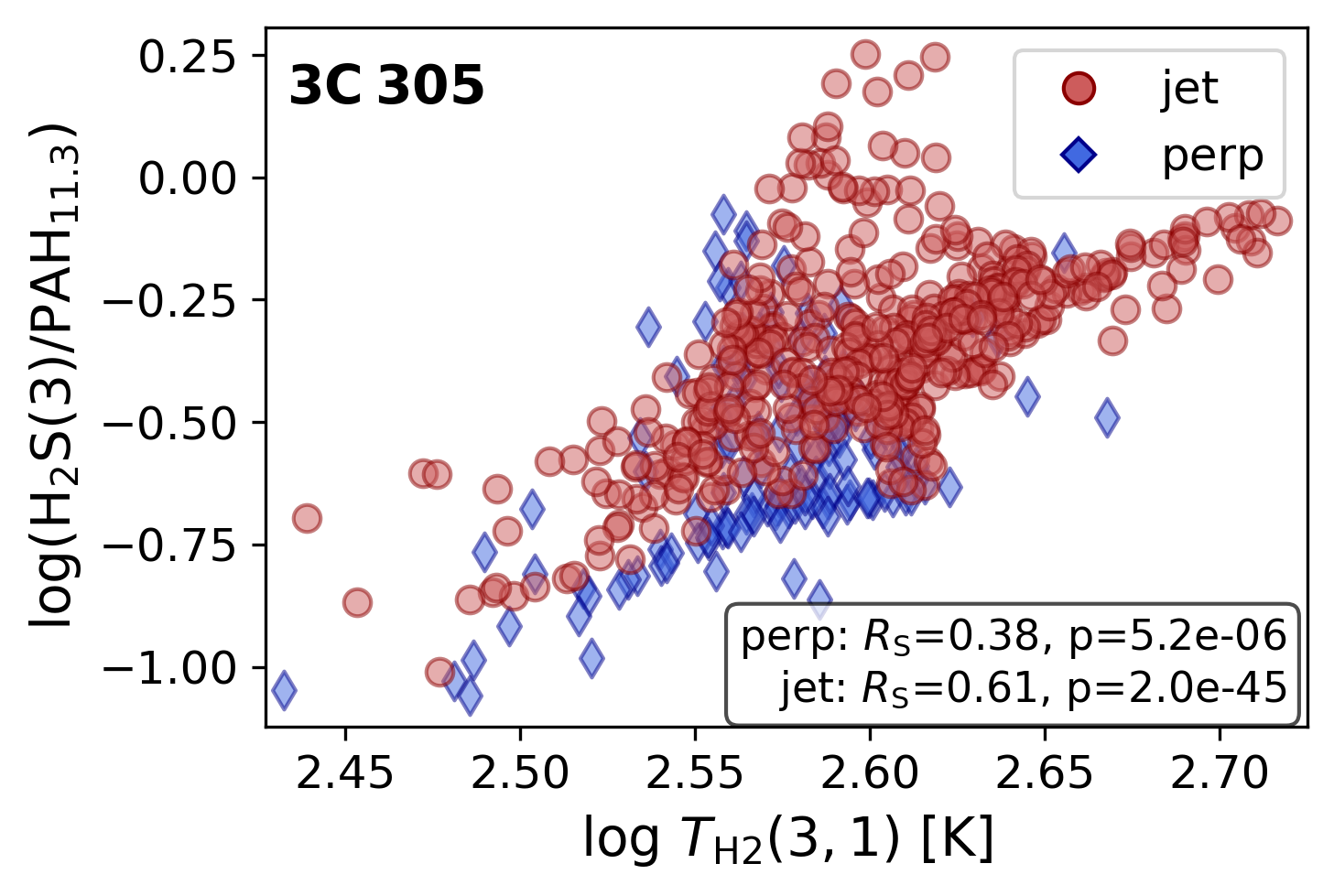} 
\includegraphics[width=0.3\textwidth]{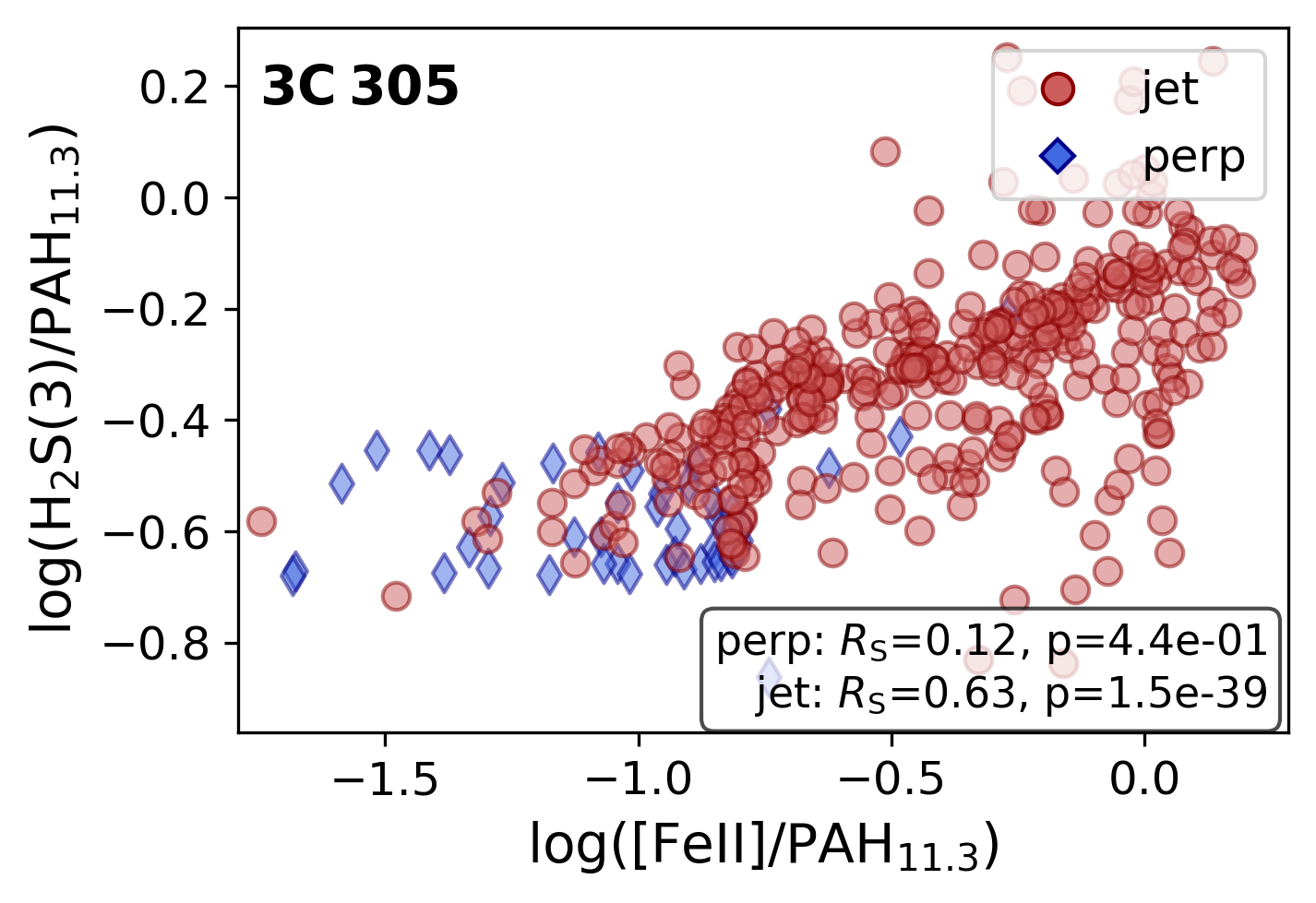} 
\includegraphics[width=0.3\textwidth]{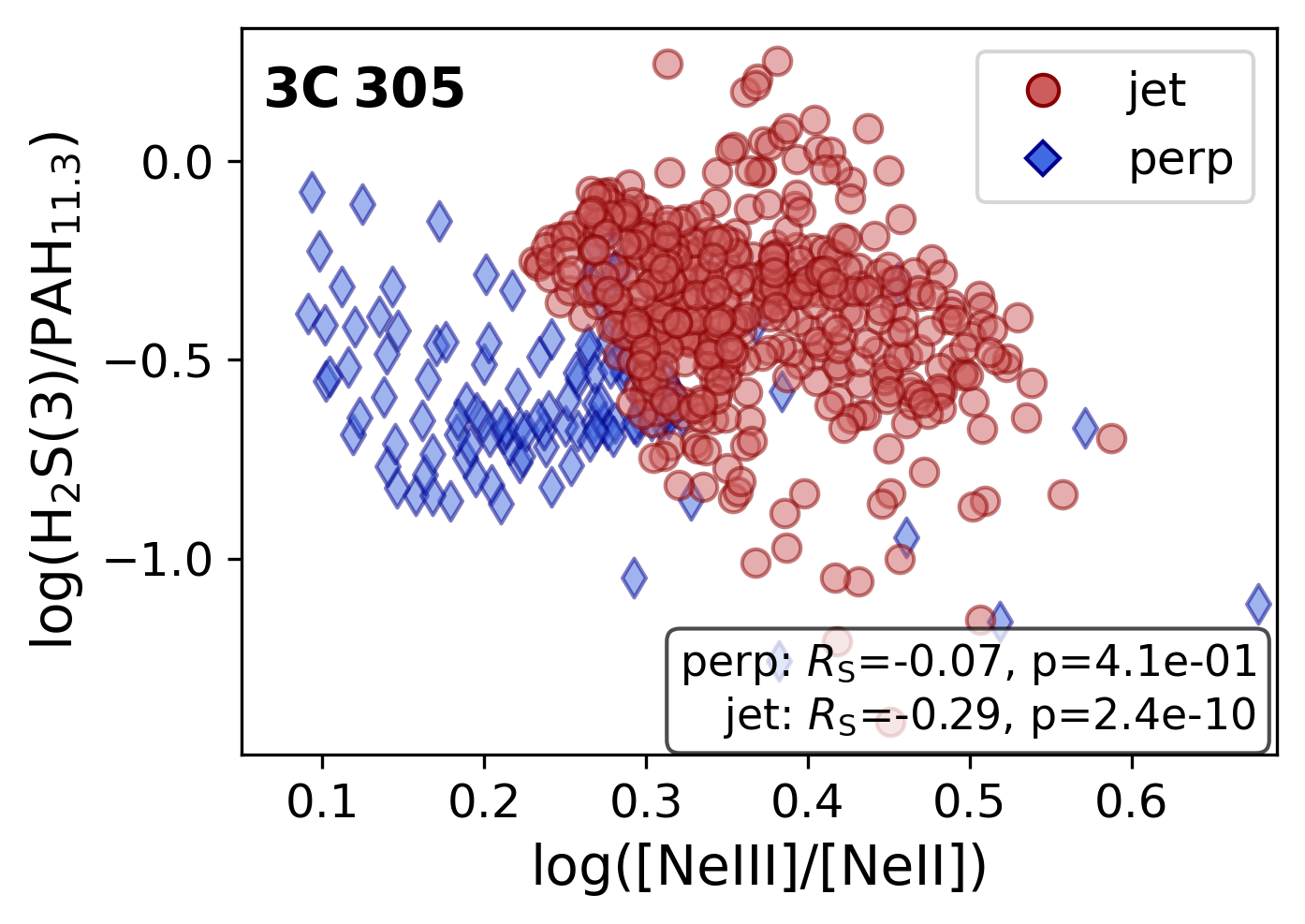}

\includegraphics[width=0.3\textwidth]{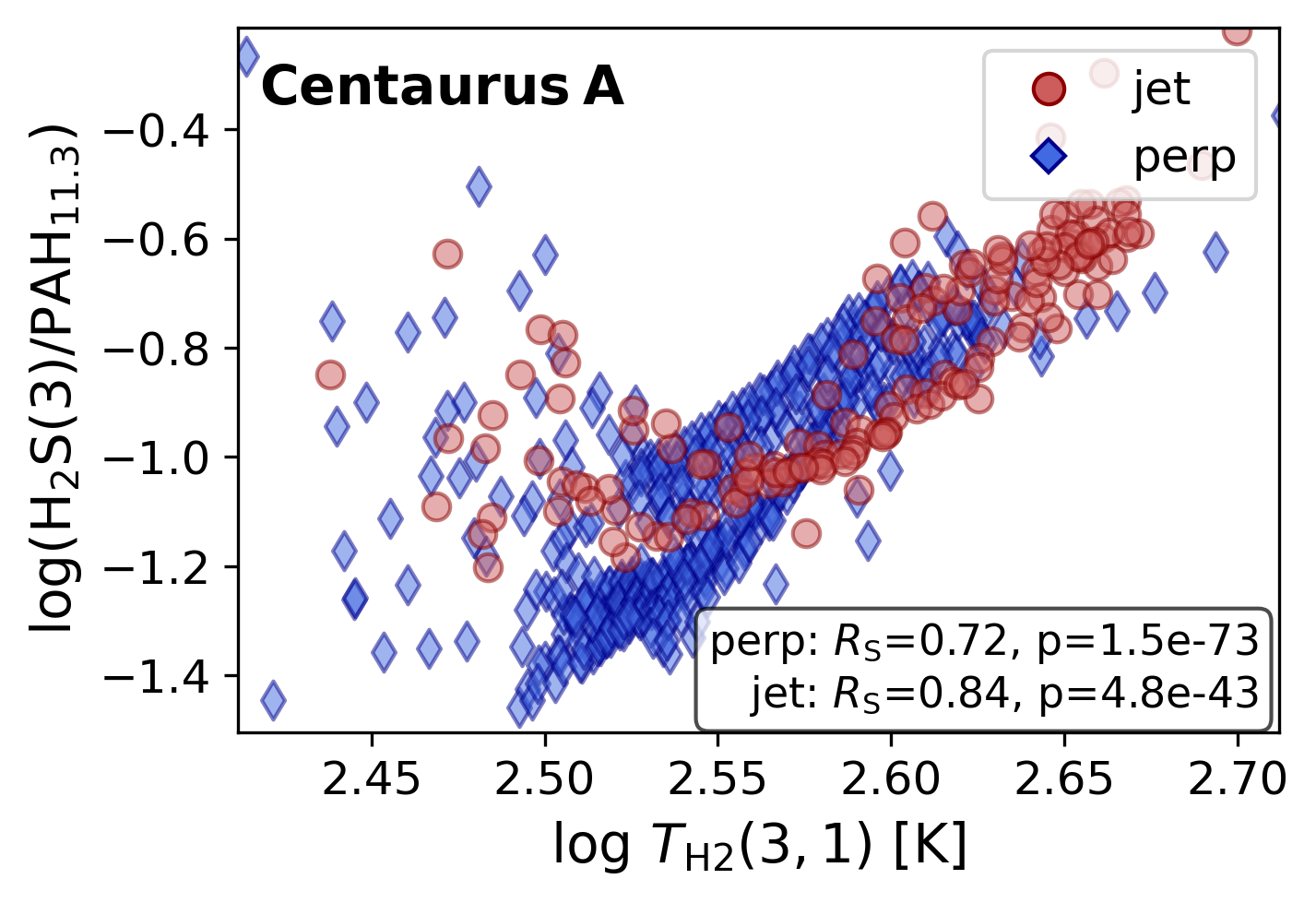} 
\includegraphics[width=0.3\textwidth]{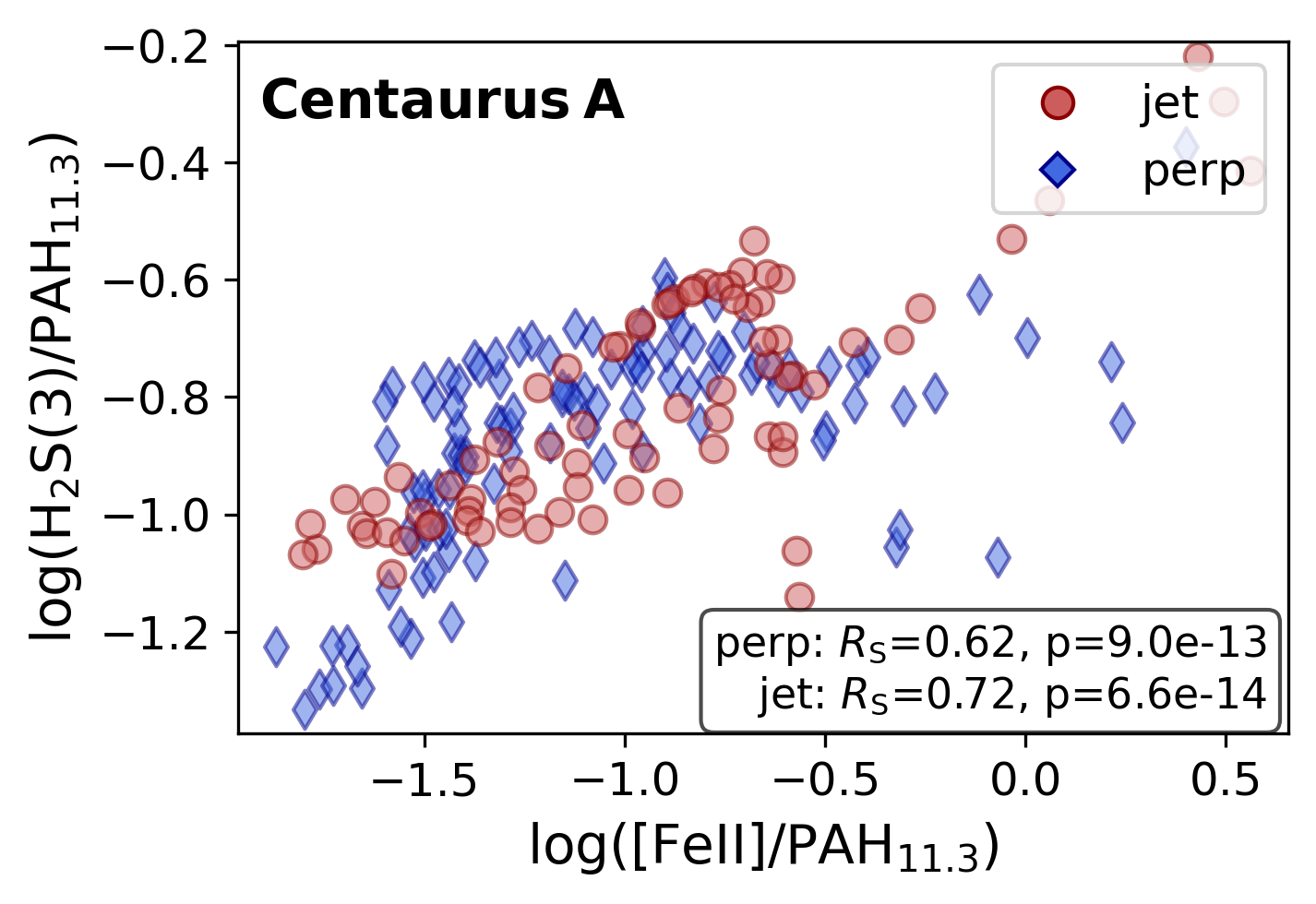} 
\includegraphics[width=0.3\textwidth]{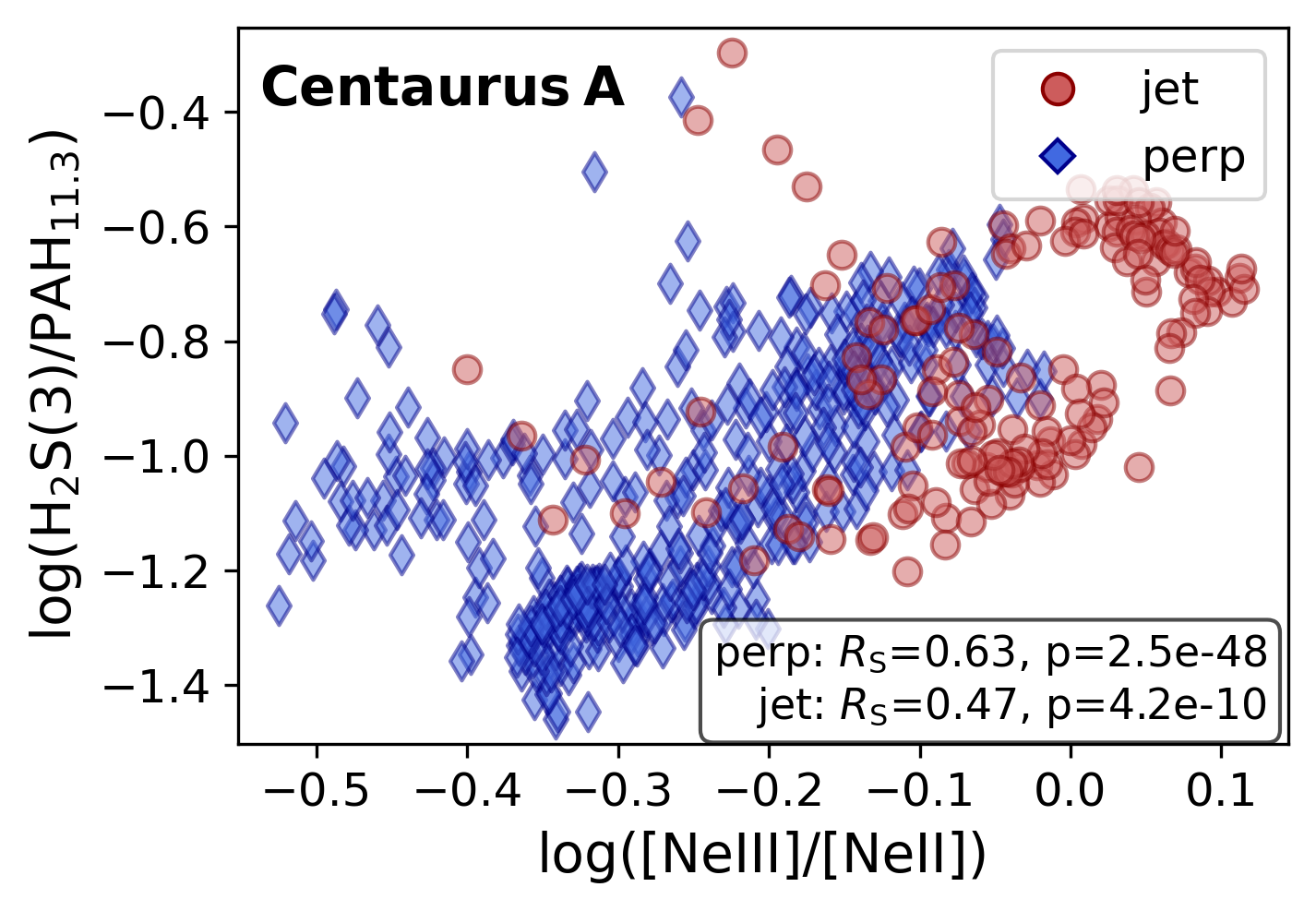}

\includegraphics[width=0.3\textwidth]{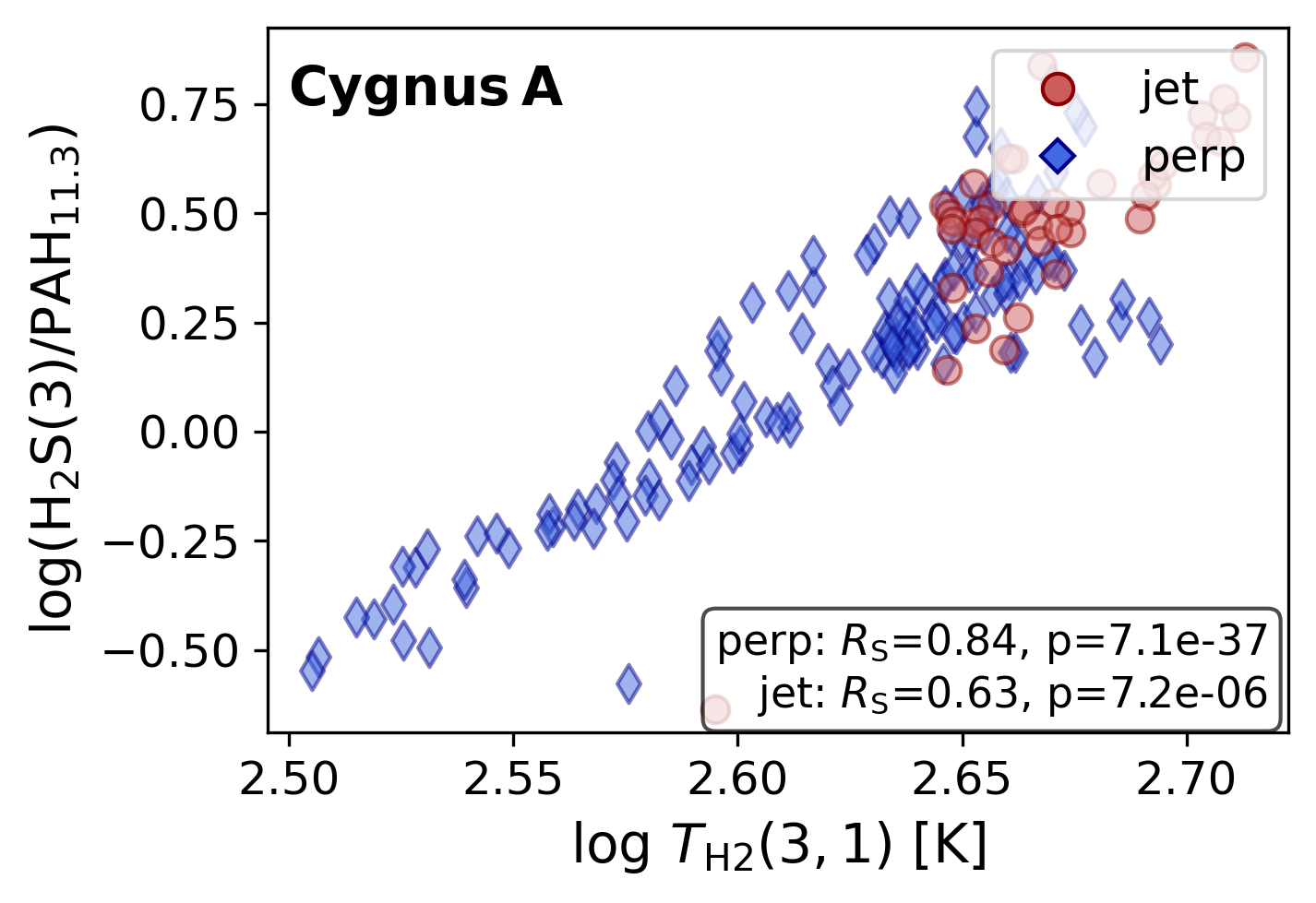} 
\includegraphics[width=0.3\textwidth]{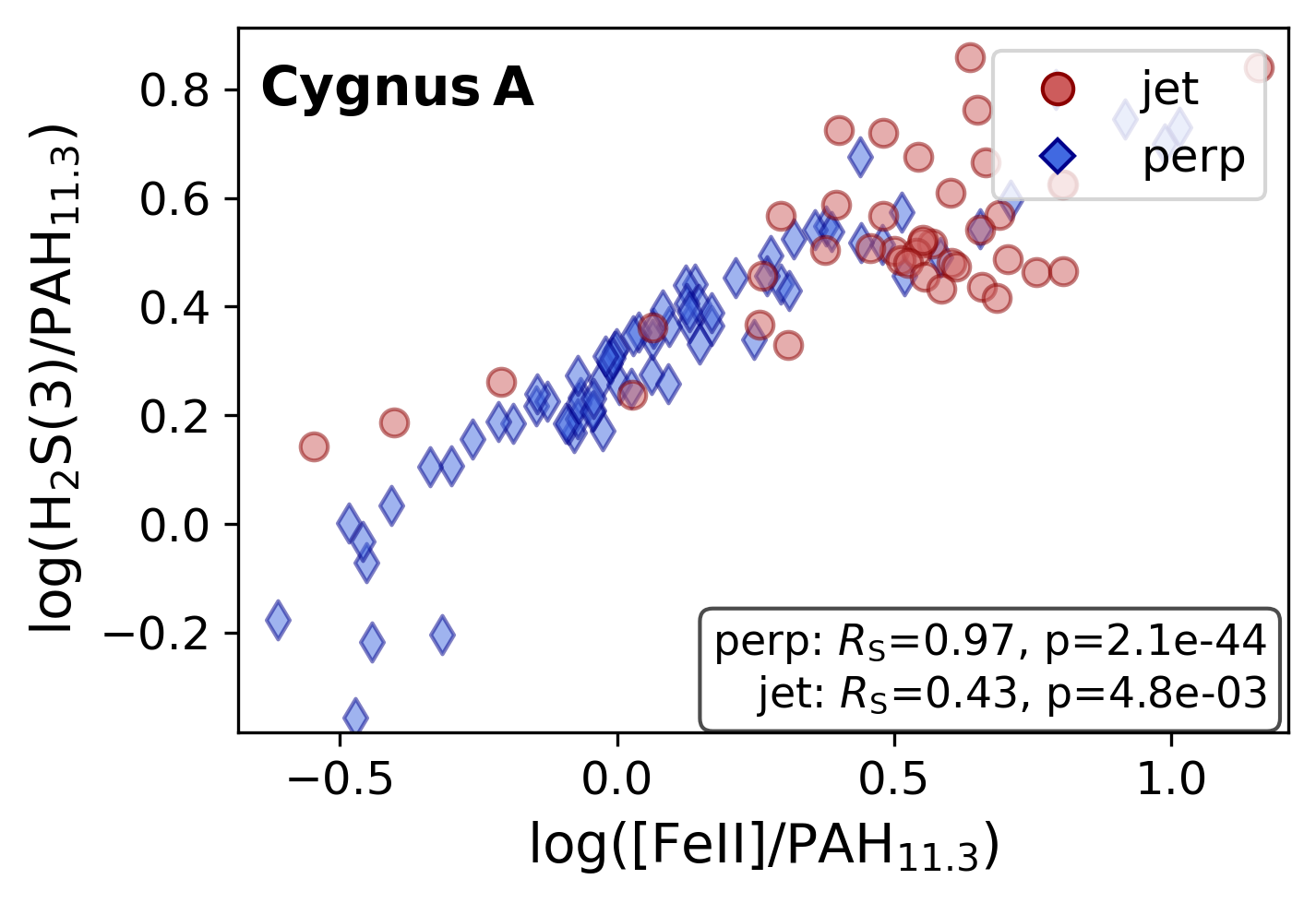} 
\includegraphics[width=0.3\textwidth]{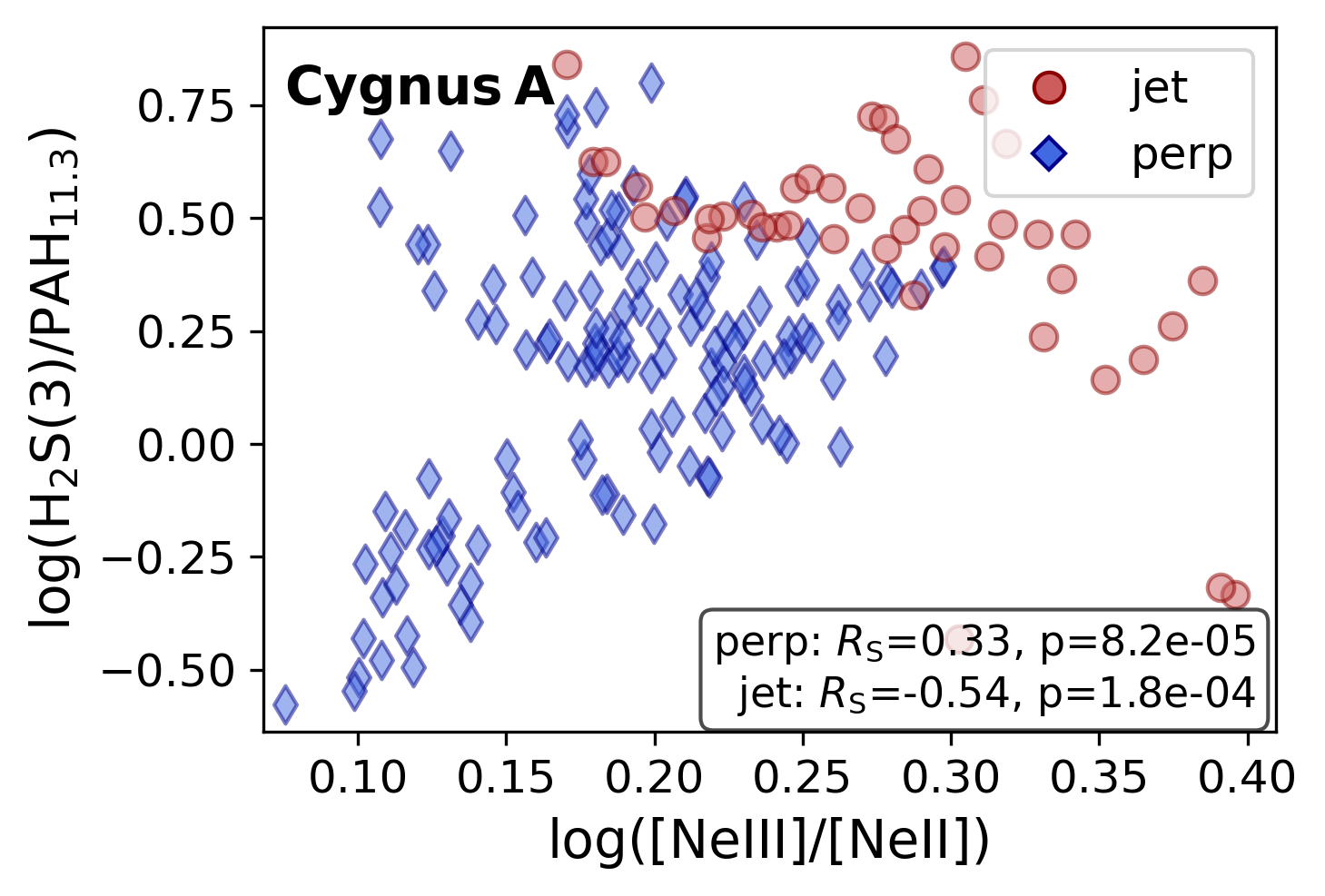}

\includegraphics[width=0.3\textwidth]{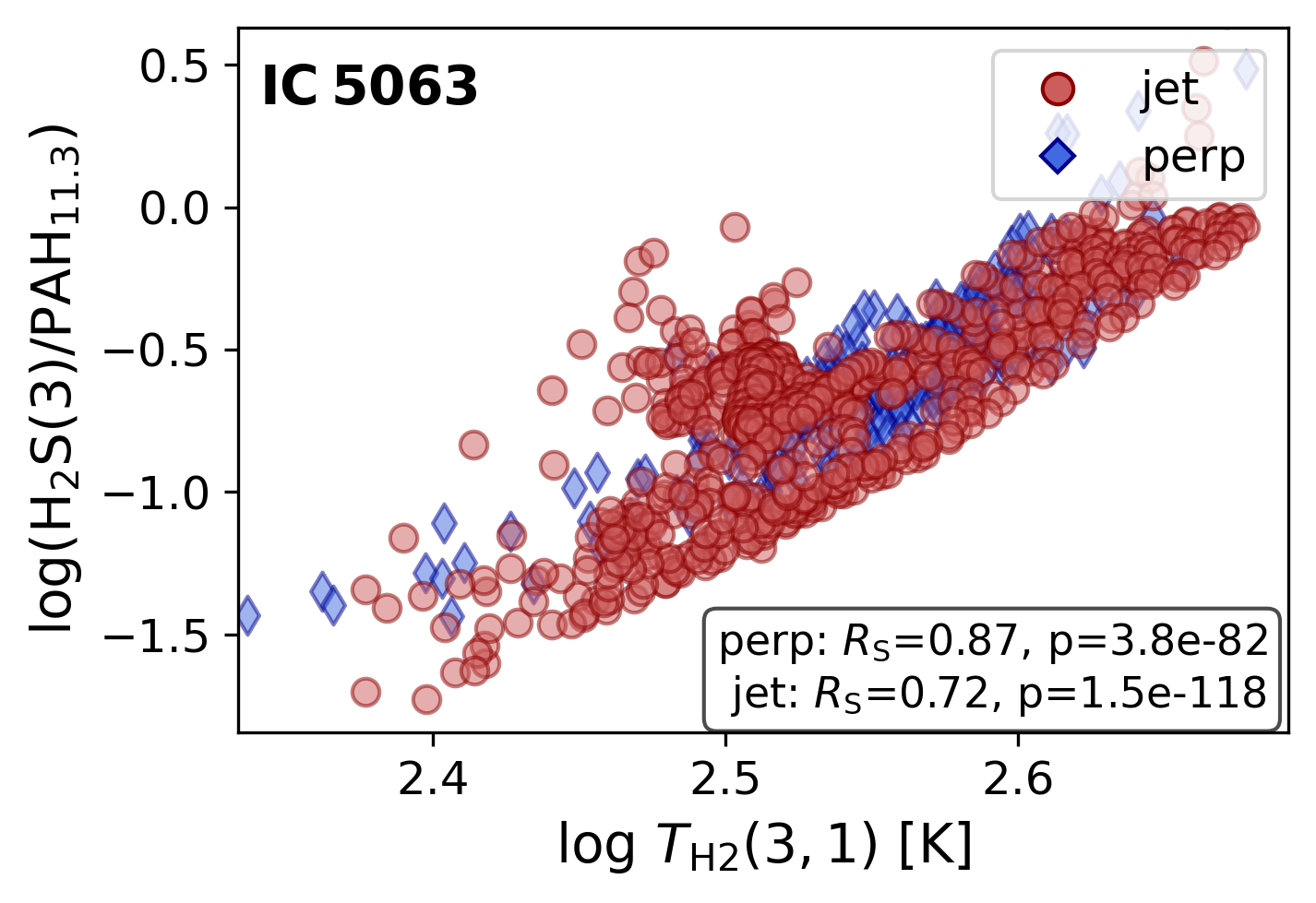} 
\includegraphics[width=0.3\textwidth]{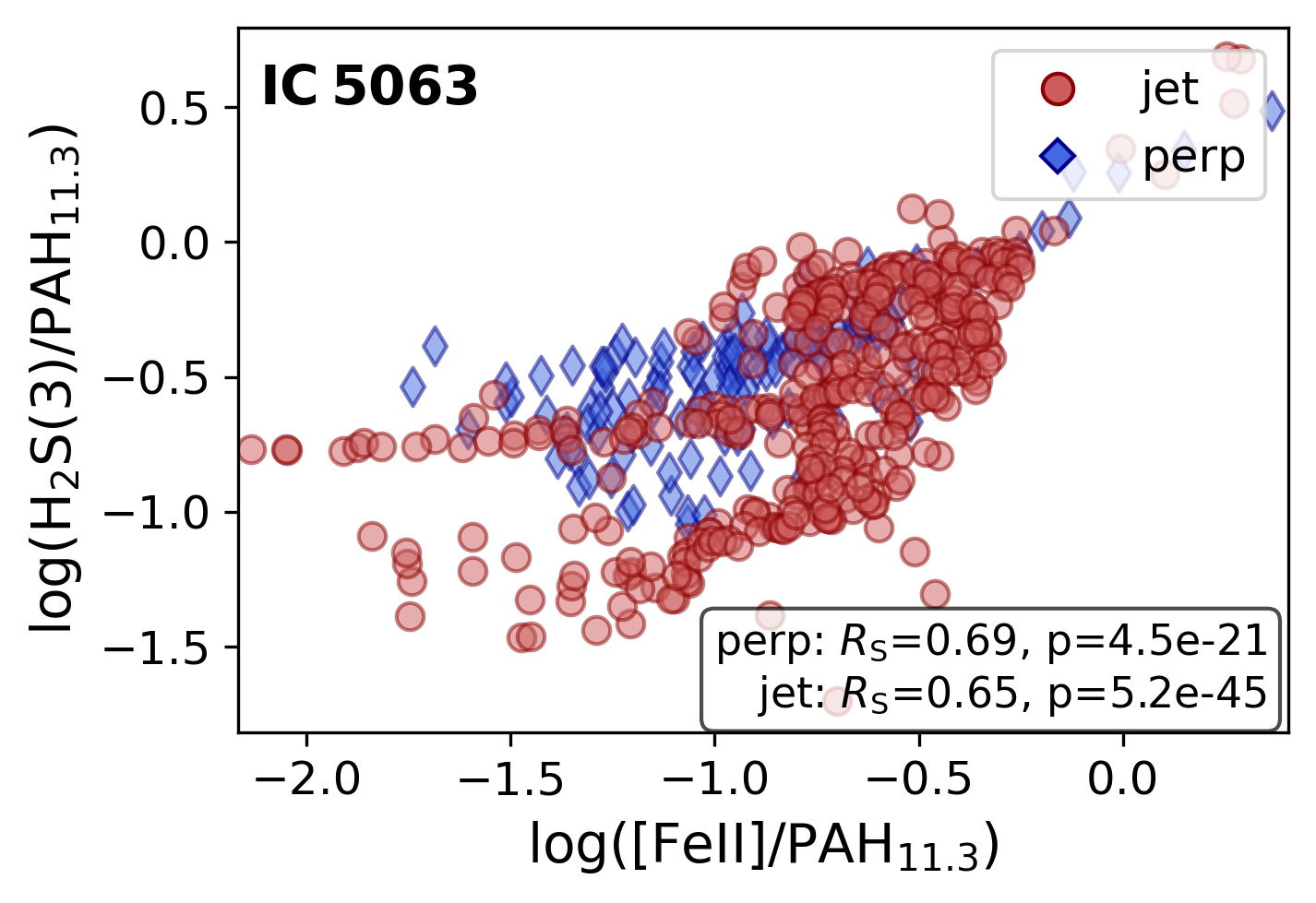} 
\includegraphics[width=0.3\textwidth]{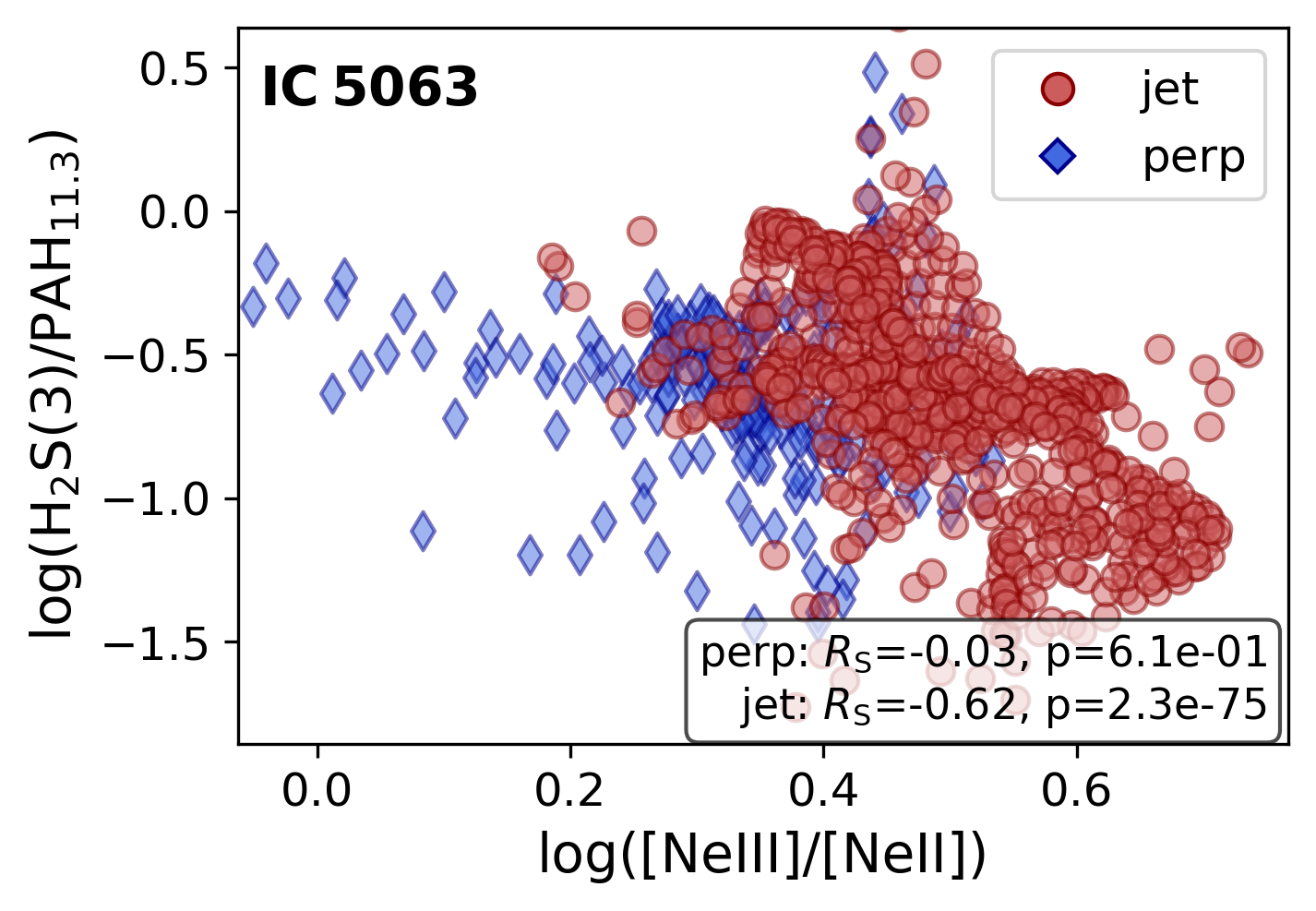}

\includegraphics[width=0.3\textwidth]{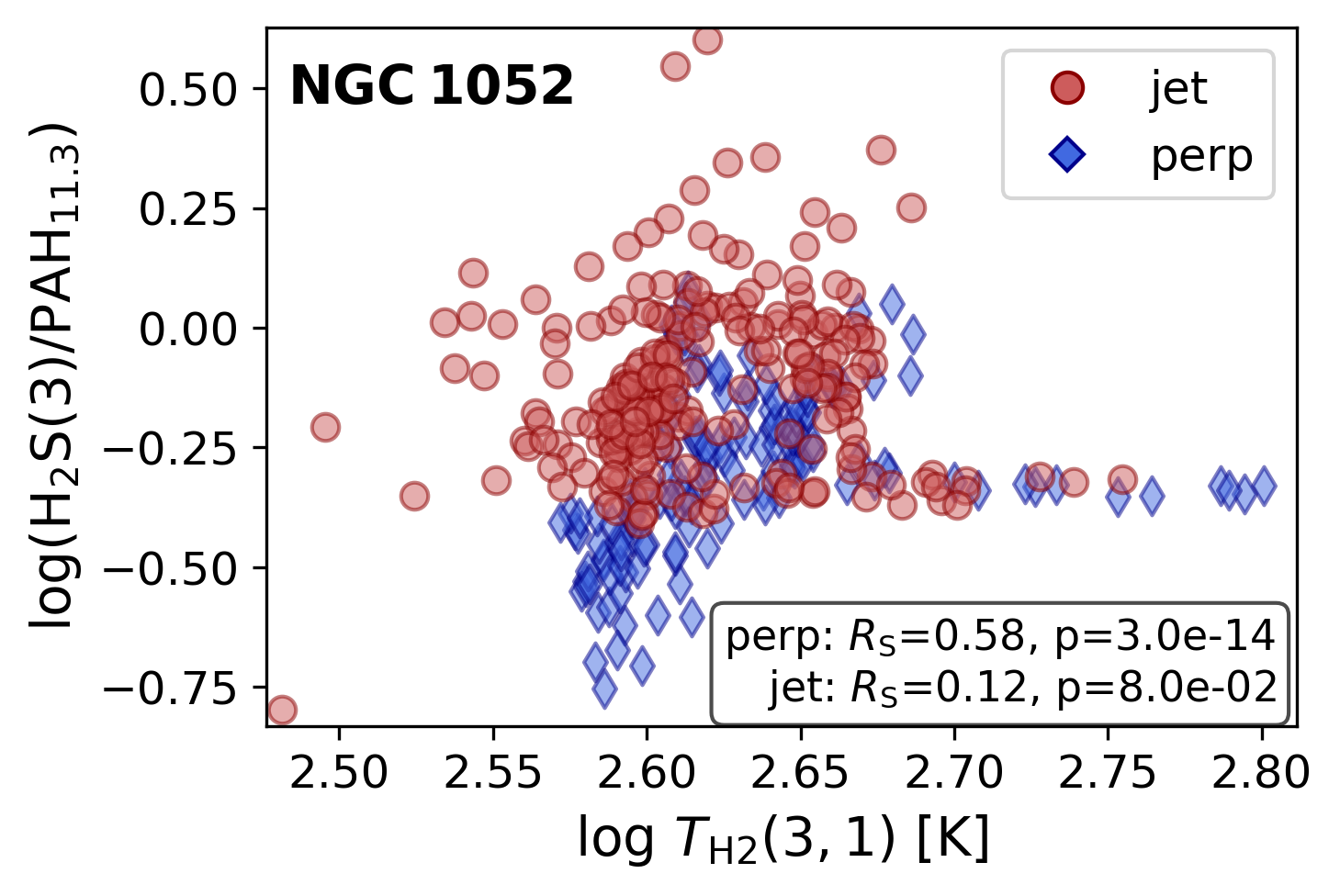} 
\includegraphics[width=0.3\textwidth]{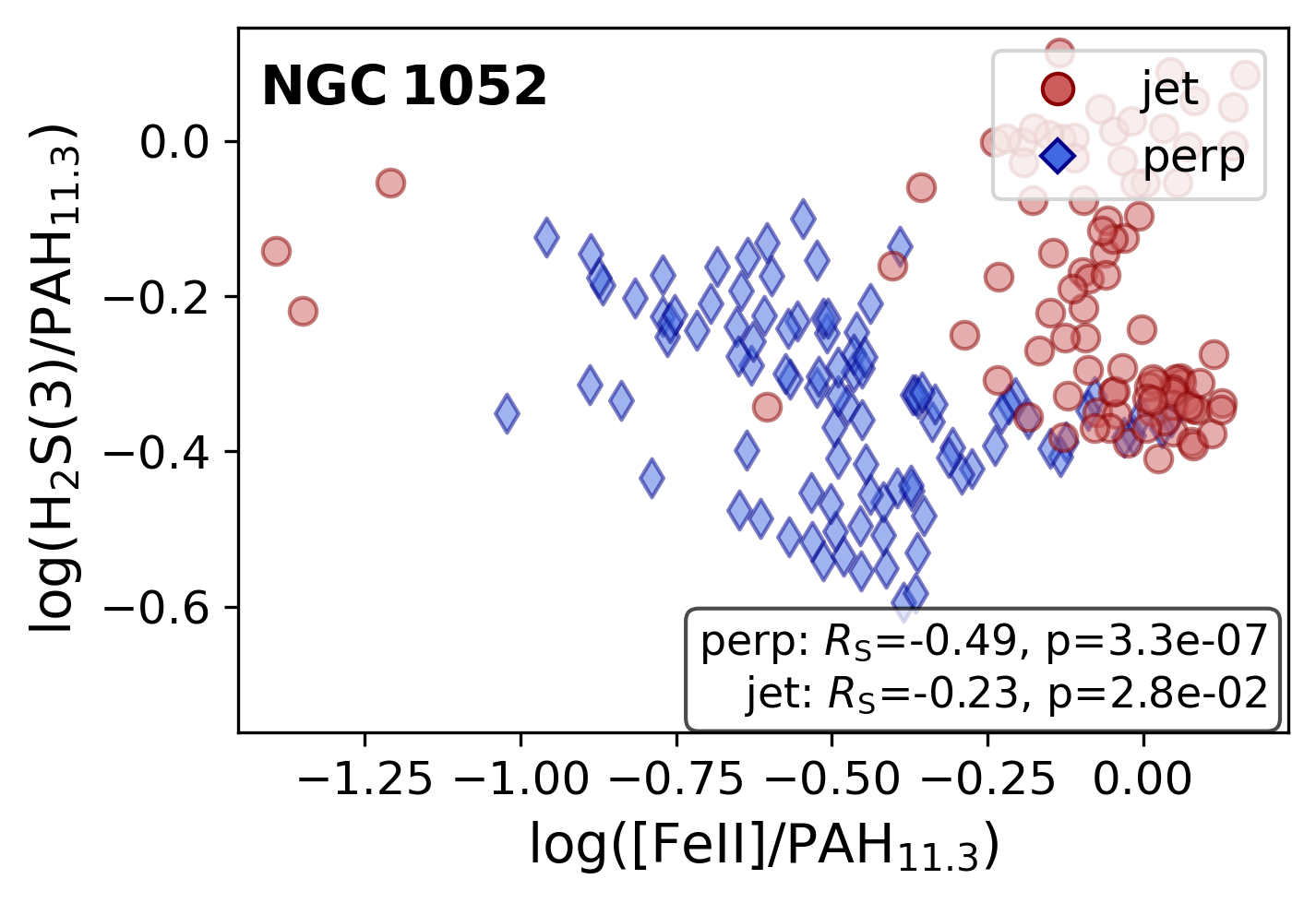} 
\includegraphics[width=0.3\textwidth]{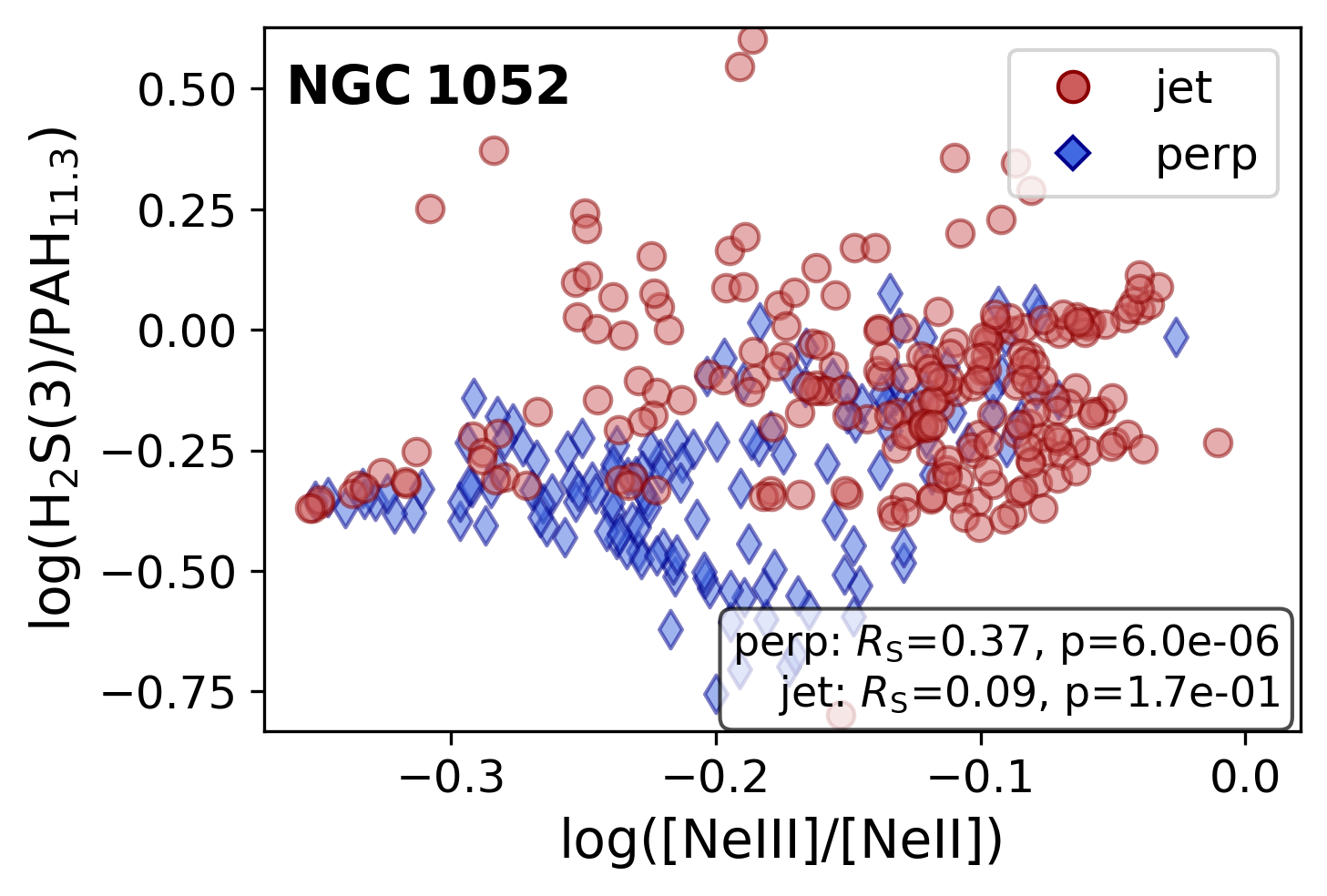}

\caption{For each galaxy, $\log\,$H$_2$\,S(3)/PAH\,11.3 versus: $\log\,T_{\rm H2}{\rm (3,1)}$ (\textit{left}), $\log\,$\,[Fe\:{\sc ii}]/PAH\,11.3 (\textit{middle}), and $\log\,$[Ne\:{\sc iii}]/[Ne\:{\sc ii}] (\textit{right}). Red circles represent the values for spaxels located within $\pm30^\circ$ of the radio jet direction, while blue diamonds correspond to spaxels located perpendicular to the jet, within $\pm30^\circ$. 
The bottom-right box in each panel shows the Spearman correlation coefficient ($R_{\rm S}$) and the p-value for each group.
 }
         \label{fig:scatter_temp}
   \end{figure*}

\subsection{Relation between H$_2$/PAH and jet orientation}

\begin{figure}
   \centering
\includegraphics[width=0.49\textwidth]{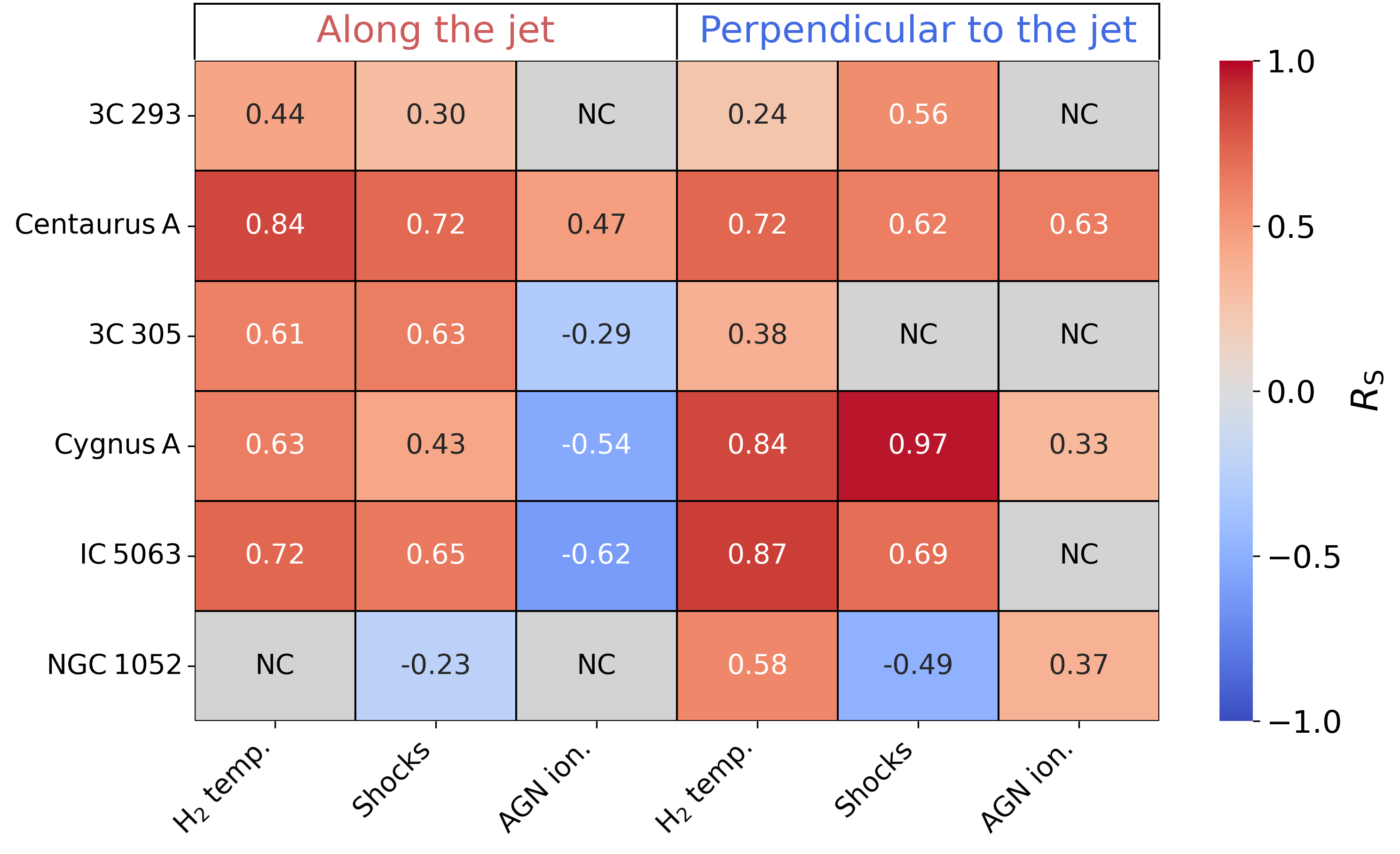} 
      \caption{Results of the Spearman correlation tests between H$_2$\,S(3)/PAH\,11.3 versus $T_{\rm H2}{\rm (3,1)}$ (H$_2$ temp.), [Fe\:{\sc ii}]/PAH\,11.3 (Shocks), and [Ne\:{\sc iii}]/[Ne\:{\sc ii}] (AGN ion.), measured both along and perpendicular to the radio axis. The numbers in each cell correspond to the measured $R_{\rm S}$ coefficients. Only correlations with statistical significance, defined as $p < 0.05$ (95\% confidence level), are shown; cases that do not satisfy this criterion are marked as NC.}
         \label{fig:correlation}
   \end{figure}

Figure~\ref{fig:scatter_temp} shows the H$_2$\,S(3)/PAH\,11.3 ratio plotted against the H$_2$ excitation temperature $T_{\rm H2}{\rm (3,1)}$ (left panels), [Fe\:{\sc ii}]/PAH\,11.3 (tracer of shocks; central panels), and  [Ne\:{\sc iii}]/[Ne\:{\sc ii}]  (tracer of the AGN ionization; right panels). Spaxels located within  $\pm30^\circ$ of the radio axis are shown as red circles, while spaxels within $\pm30^\circ$ of the axis perpendicular to the jet are shown as blue diamonds, with only the nuclear spaxel overlapping. We tested correlations using Spearman's rank coefficient, which are listed in the bottom-right corner of each panel and summarized in Fig.~\ref{fig:correlation}. 

Along the jet direction, we find that the H$_2$\,S(3)/PAH\,11.3 ratio correlates with both $T_{\rm H2}{\rm (3,1)}$ and [Fe\,{\sc ii}]/PAH\,11.3 for all galaxies, with the exception of NGC\,1052, which shows no correlation with $T_{\rm H2}{\rm (3,1)}$ and a weak anticorrelation with [Fe\,{\sc ii}]/PAH\,11.3. On the other hand, the H$_2$\,S(3)/PAH\,11.3 ratio and [Ne\,{\sc iii}]/[Ne\,{\sc ii}] are anticorrelated in three galaxies (3C\:305, Cygnus\:A, and IC\:5063). Two galaxies (3C\:293 and NGC\:1052) show no significant correlation, while Centaurus\:A exhibits a positive correlation.

Perpendicular to the jet orientation, all galaxies show correlations between H$_2$\,S(3)/PAH\,11.3 and $T_{\rm H2}{\rm (3,1)}$. For Cygnus\,A, IC\,5063, and NGC\,1052, these correlations are stronger than those measured along the jet, while the remaining galaxies show weaker correlations compared to their corresponding jet-direction values. In addition, four galaxies show correlations between H$_2$\,S(3)/PAH\,11.3 and [Fe\,{\sc ii}]/PAH\,11.3; the exceptions are 3C\,305, which shows no correlation, and NGC\,1052, which displays an anticorrelation. Finally, correlations between the H$_2$\,S(3)/PAH\,11.3 ratio and [Ne\,{\sc iii}]/[Ne\,{\sc ii}] are found in three galaxies (Centaurus\,A, Cygnus\,A, and NGC\,1052), while the remaining three (3C\,293, 3C\,305, and IC\,5063) show no significant correlation.

\begin{figure}
   \centering
\includegraphics[width=0.45\textwidth]{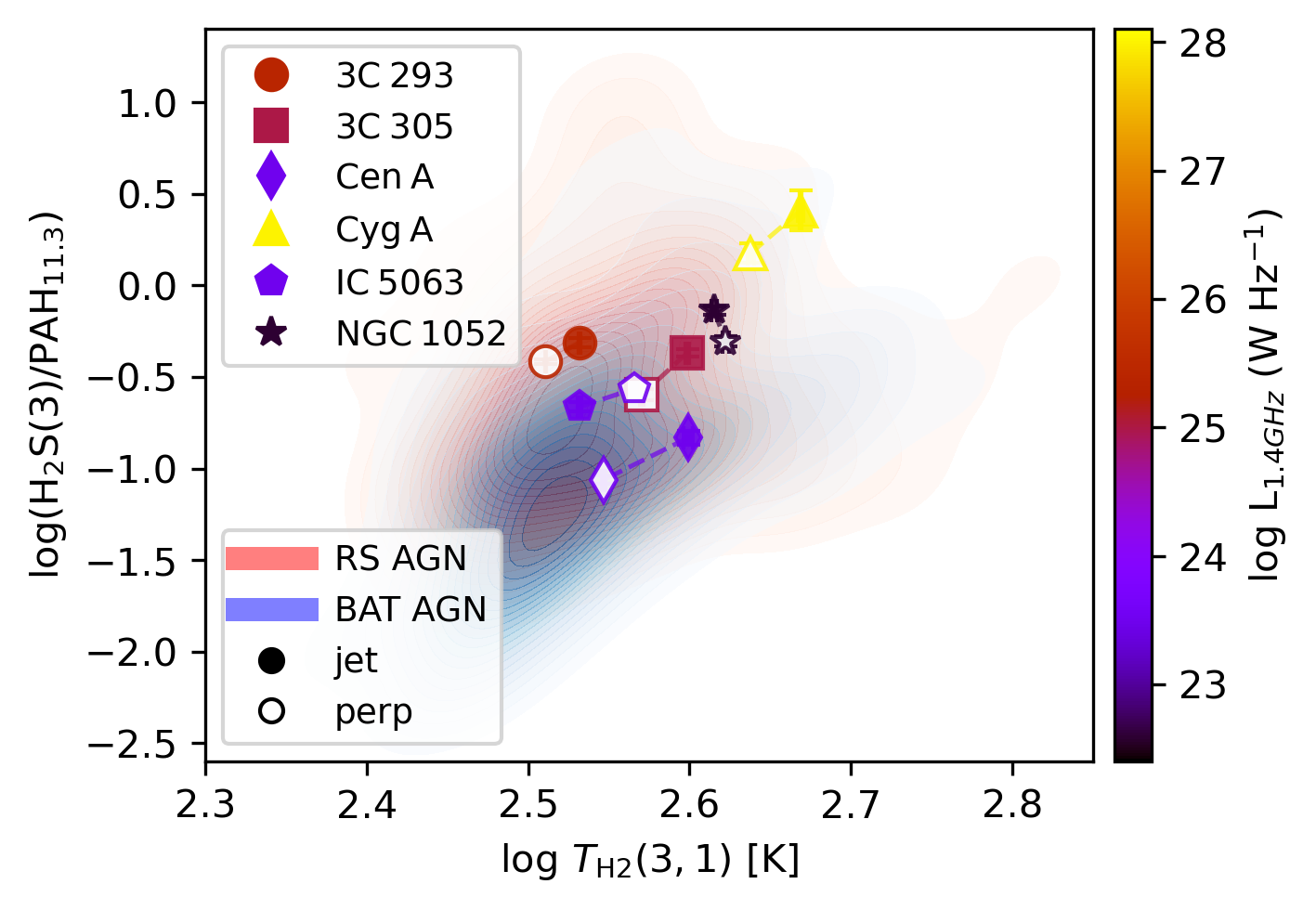} 
\includegraphics[width=0.45\textwidth]{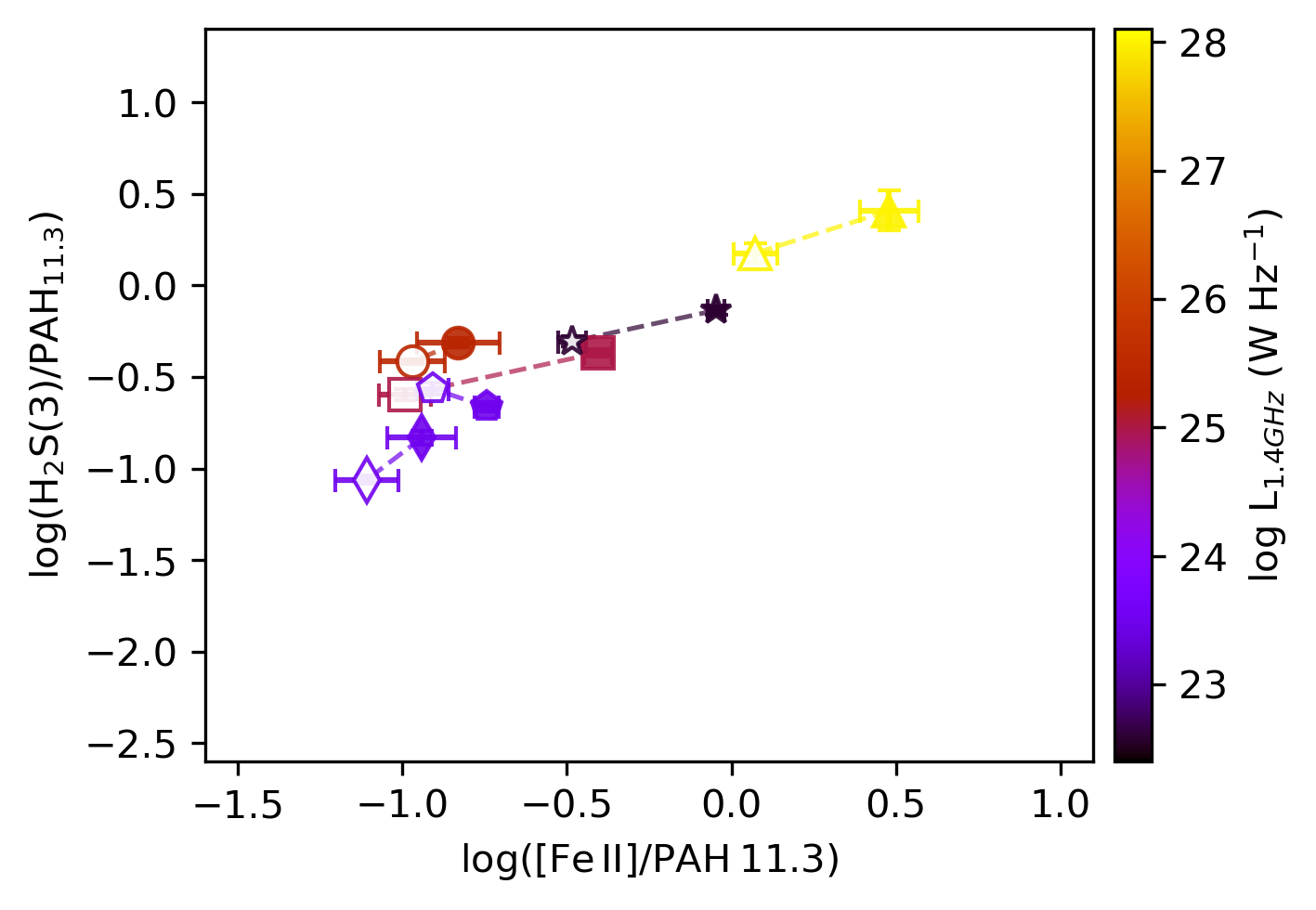}
\includegraphics[width=0.45\textwidth]{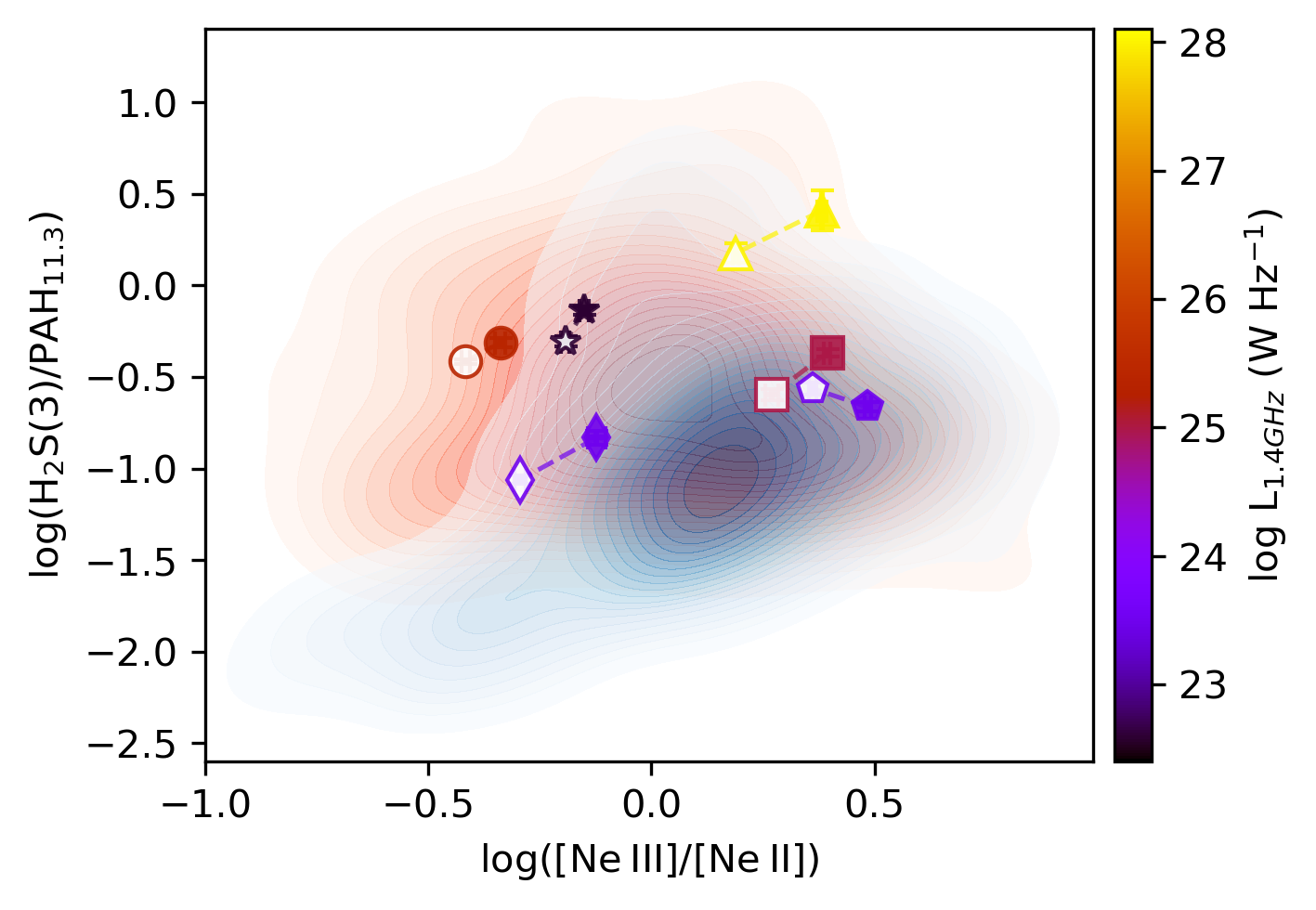} 
      \caption{ $\langle\log\,$H$_2$\,S(3)/PAH\,11.3$\rangle$ versus: $\langle \log\,T_{\rm H2}{\rm (3,1)}\rangle$ (\textit{top}), $\langle\log\,$[Fe\,{\sc ii}]/PAH\,11.3$\rangle$ (\textit{center}), and $\langle \log\,$[Ne\,{\sc iii}]/[Ne\,{\sc ii}]$\rangle$ (\textit{bottom}), showing mean logarithmic line ratios measured along (filled symbols) and perpendicular to (open symbols) the jet directions. Error bars indicate the standard errors of the mean values. Points are color-coded by the radio luminosity, as indicated by the colorbar. Thin dashed lines connect the two points for each galaxy. The density plots are based on \textit{Spitzer} measurements for the BAT AGN and RS AGN samples, as defined in Sect.~\ref{sec:ew}.}
         \label{fig:scatter_mean}
   \end{figure}

Figure~\ref{fig:scatter_mean} shows the mean logarithmic values of the line ratios ($R$), computed as the average of $\log(R)$ considering spaxels along the radio jet (filled symbols) and perpendicular to the jet (open symbols). The top panel displays $\log(\mathrm{H_2\,S(3)/PAH\,11.3})$ versus $\log T_{\mathrm{H_2}}(3,1)$, the middle panel shows $\log([\mathrm{Fe\,II}]/\mathrm{PAH\,11.3})$, and the bottom panel $\log([\mathrm{Ne\,III}]/[\mathrm{Ne\,II}])$. The error bars show the standard error of the mean, computed considering the effective number of independent resolution elements, and in some cases the corresponding uncertainties are smaller than the size of the markers. Density plots for the BAT and RS AGN samples (see Sect.~\ref{sec:ew}), based on \textit{Spitzer} measurements \citep{Spoon22}, are presented in the top and bottom panels. The middle panel does not include such density contours because the [Fe\,{\sc ii}]\,5.34\,$\mu$m line is not covered in the measurements reported by \citet{Spoon22}. 

The comparison with \textit{Spitzer}-based measurements shows that the H$_2$ excitation temperatures derived from the MIRI MRS observations are, on average, higher than those derived from the integrated \textit{Spitzer} spectra. Similarly, the mean logarithmic ratio, $\langle\log($H$_2$\,S(3)/PAH\,11.3$)\rangle$, is also slightly higher than the \textit{Spitzer}-based values. This difference is likely due to the larger aperture of the \textit{Spitzer} spectra, which may include emission from extra-nuclear star-forming regions that exhibit lower excitation temperatures and H$_2$/PAH ratios \citep{Petric18,lambrides19,rogemar25_jwst}. Excluding NGC\:1052, which has the lowest radio power in our sample, both the $\mathrm{H_2\,S(3)/PAH\,11.3}$ and $[\mathrm{Fe\,II}]/\mathrm{PAH\,11.3}$ ratios, as well as the excitation temperature, increase with radio power.  On the other hand, the $[\mathrm{Ne\,III}]/[\mathrm{Ne\,II}]$ ratios measured with JWST fall within the range of values obtained from \textit{Spitzer} observations. 
No clear trend is observed between this line ratio and radio power.

The comparison of the mean H$_2$ excitation temperatures and line ratios along and perpendicular to the radio jet shows that the H$_2$/PAH ratio is higher along the jet for all galaxies except IC\,5063; the H$_2$ excitation temperature is higher along the jet for four galaxies, but lower for IC\,5063 and NGC\,1052; the $[\mathrm{Fe\,II}]/\mathrm{PAH\,11.3}$ ratio is larger along the jet for all sources; and the $[\mathrm{Ne\,III}]/[\mathrm{Ne\,II}]$ ratio is higher along the jet for all galaxies compared to the perpendicular direction.

The interpretation and discussion of the results described in this section are presented in Sects.~\ref{sec:disc-lw} and \ref{sec:disc-lr}.

\subsection{Mean gas velocity dispersion across distinct regions}

   \begin{figure}
   \centering
\includegraphics[width=0.49\textwidth]{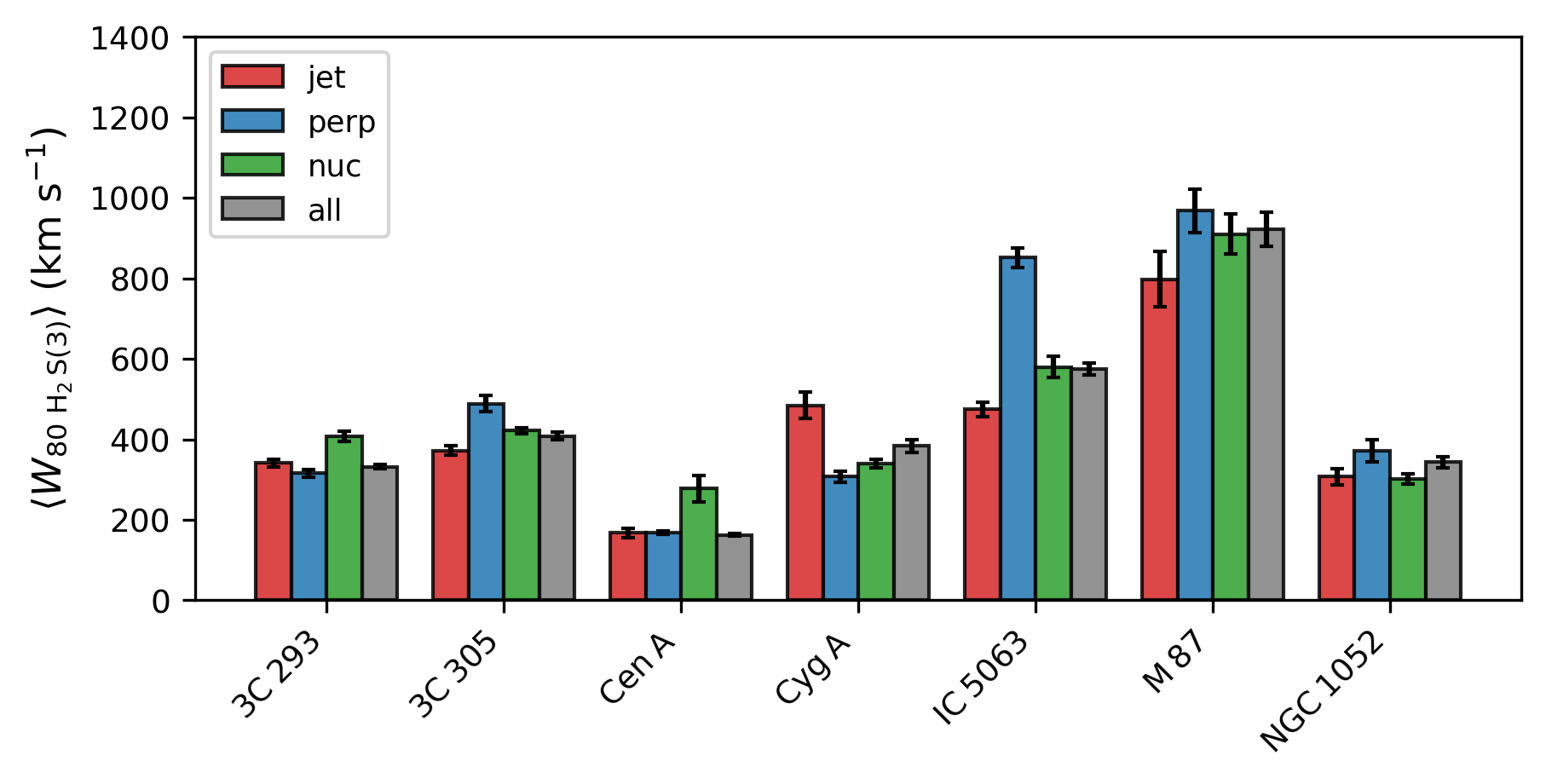} 
\includegraphics[width=0.49\textwidth]{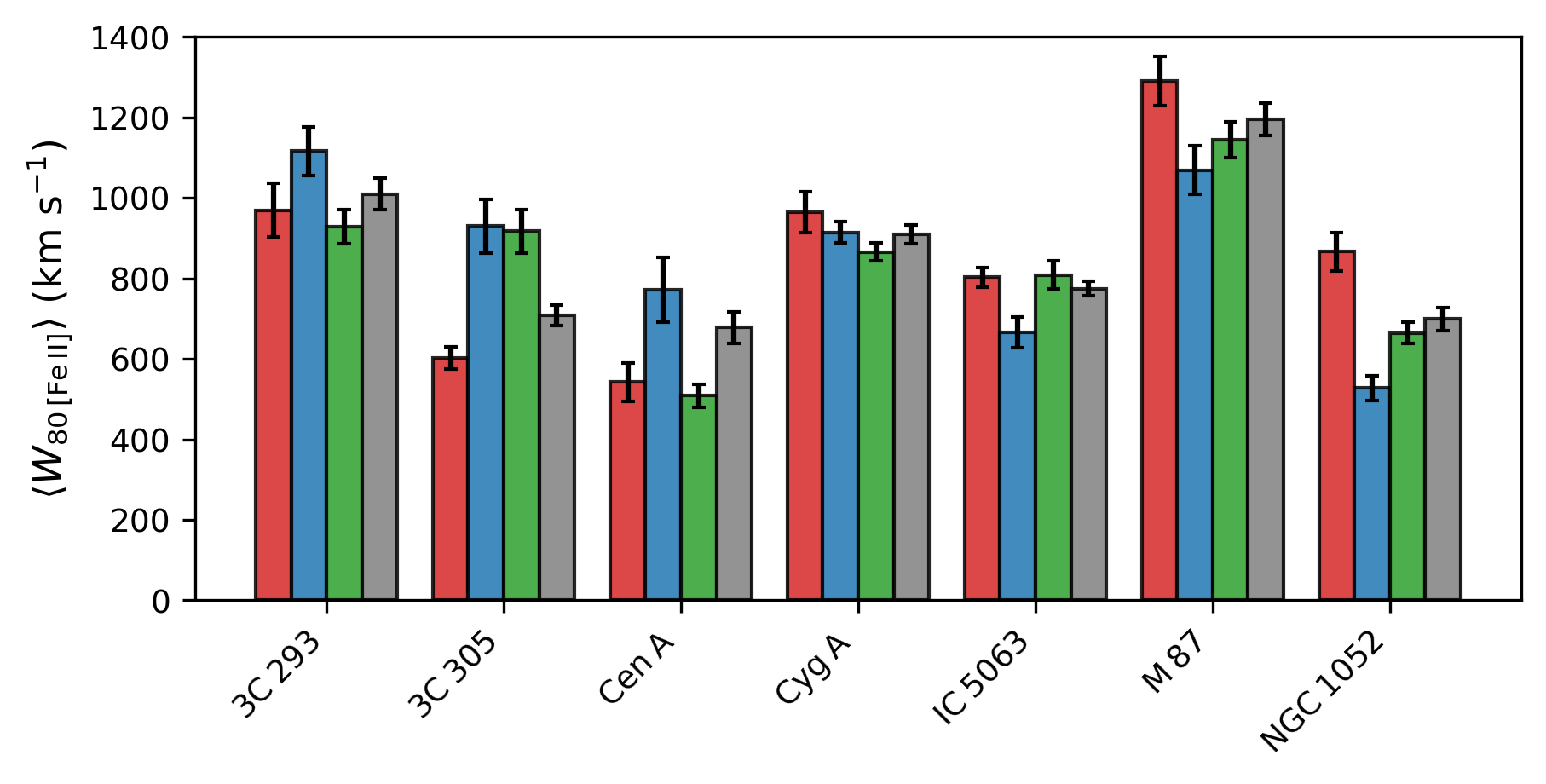} 
\includegraphics[width=0.49\textwidth]{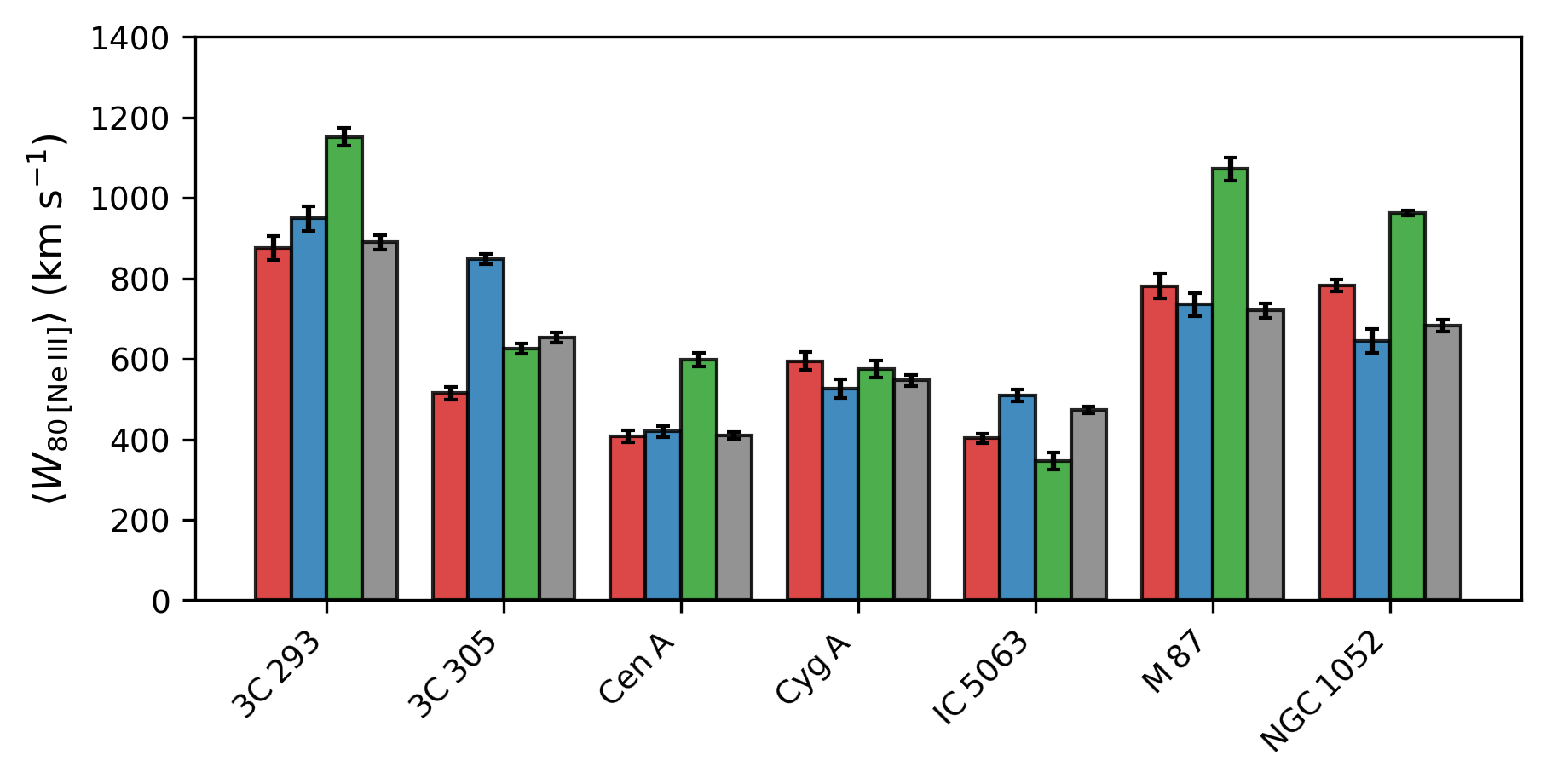} 
  \caption{Mean W$_{80}$ values for the H$_2$ S(3) (\textit{top}), [Fe\,{\sc ii}] (\textit{middle}), and [Ne\,{\sc iii}] (\textit{bottom}) lines, computed using spaxels selected along the radio-jet direction ($\pm30^\circ$), perpendicular to the jet ($\pm30^\circ$), within the nuclear region ($r<1$ arcsec), and using all spaxels for each galaxy. The error bars correspond to the standard error of the mean, computed considering the effective number of independent resolution elements contributing to each measurement.}
         \label{fig:histW80}
   \end{figure}

Figure~\ref{fig:histW80} presents the mean $W_{\rm 80}$ values for H$_2$\:S(3) (top), [Fe\:{\sc ii}] (middle), and [Ne\:{\sc iii}] (bottom) measured in four different regions: along the jet (jet), perpendicular to the jet (perp), within the inner 1\,arcsec radius (nuc), and considering all spaxels (all). The lowest $W_{\rm 80}$ values are observed for the molecular gas, except for IC\,5063 and M\,87, which show slightly lower values for [Ne\:{\sc iii}].  The highest velocity dispersion is observed for [Fe\:{\sc ii}], with $W_{\rm 80}$ values exceeding 500\,km\,s$^{-1}$ and reaching up to $\sim$1200\,km\,s$^{-1}$ for M\,87. The highest $W_{\rm 80}$ values within the nuclear aperture are observed for only two galaxies in H$_2$. None of the galaxies exhibit their highest [Fe\,{\sc ii}] $W_{\rm 80}$ values in the nucleus, whereas four galaxies show the highest [Ne\,{\sc iii}] $W_{\rm 80}$ values within the nuclear aperture. 

The comparison of the mean $W_{\rm 80}$ values along and perpendicular to the radio jet reveals enhanced [Fe\,{\sc ii}] gas turbulence perpendicular to the jet for 3C\,293, 3C\,305, and Centaurus\,A, while the remaining four galaxies show higher $W_{\rm 80}$ values along the jet direction. For the [Ne\,{\sc iii}] emission, the same behavior is observed, with enhanced velocity dispersion also present in IC\,5063. Figure~\ref{fig:maps_IC5063} shows that IC\,5063 exhibits increased velocity dispersion perpendicular to the jet, but not centered on the galaxy nucleus, consistent with a biconical outflow as discussed by \citet{Dasyra24}. For the molecular gas, 3C\,293 and Cygnus\,A exhibit higher $W_{\rm 80}$ values along the jet direction; 3C\,305, IC\,5063, NGC\,1052, and M\,87 show higher values perpendicular to the jet, while Centaurus\,A displays similar values along and perpendicular to the radio jet. Thus, considering both the ionized and molecular gas phases, all galaxies except Cygnus\,A exhibit enhanced gas turbulence in regions perpendicular to the radio jet. We further discuss the origin of this turbulence in Sect.~\ref{sec:disc-lw}.

\section{Discussion}\label{sec:disc}

\subsection{Enhanced gas turbulence perpendicular to radio jets}\label{sec:disc-lw}

Among the seven galaxies studied here, six show enhanced gas velocity dispersion perpendicular to the jet axis in at least one gas phase. The only exception is Cygnus\:A, the most powerful radio source in our sample. Similar enhanced line widths perpendicular to the radio jet have been reported for several other galaxies, predominantly in the ionized gas \citep[e.g.,][]{Couto13,Couto17,Schnorr14,allan16,Lena15,Freitas18,Finlez18,Shin19,Feruglio20,venturi21,Girdhar22,Peralta23,Ulivi24,Speranza24,Hermosa24a}, but also in the hot molecular phase traced by near-IR H$_2$ emission \citep{Diniz15,rogemar14_n5929_let,rogemar_n5929}, as well as in the cold molecular gas traced by CO emission features \citep{Murthy19,Girdhar22,Ramos-Almeida22,Audibert23,Audibert25,Oosterloo25,Bruno26}. Based on JWST/MIRI MRS observations of the galaxies in our sample, enhanced gas velocity dispersion perpendicular to the jet has been reported in the ionized phase for Centaurus\,A by \citet{Alonso-Herrero24}, and in both the warm molecular and ionized gas for IC\,5063 by \citet{Dasyra24}. In addition, enhanced velocity dispersion perpendicular to the radio jet/outflow direction has been reported for other nearby AGN hosts based on MIRI/MRS data \citep{Davies24,Zhang24,Garcia-Bernete24}. 
 Cold molecular gas observations, traced by CO emission, indicate that all galaxies host disks or dust lanes that are significantly inclined with respect to the plane of the sky (with the exception of M\,87; see Table~\ref{tab:sample}).

Enhanced gas velocity dispersion measured in regions perpendicular to the radio jet axis is consistently associated with shock-excited emission, as evidenced by the systematically elevated optical line ratios, in particular the [O\,{\sc i}]$\lambda6300$/H$\alpha$ ratio, together with increased electron temperatures; these combined diagnostics reveal the presence of shock heating and are interpreted as the result of the interaction between wide-angle outflows and the surrounding ISM \citep{rogemar21_te,venturi21}.  These observations are consistent with numerical simulations that follow the interaction of relativistic jets with the galactic disk, particularly in scenarios where the jet is launched with only a small inclination ($\lesssim45^\circ$) relative to the disk plane \citep{Mukherjee18,Mukherjee25}. During its propagation through the disk, the jet progressively loses momentum as it interacts with dense gas clouds, becoming decelerated and partially deflected. This coupling with the disk gas inflates an over-pressurized, sub-relativistic wide-angle outflow that emerges approximately perpendicular to the disk plane \citep{Mukherjee18,Meenakshi22,Davies24}. In this scenario, turbulence is enhanced perpendicular to the jet, while motions along the jet axis are dominated by bulk outflow \citep{Mukherjee18}.
 Thus, the relative orientation between the radio jet and the molecular disk provides a useful indicator of jet--disk coupling. According to Table~\ref{tab:sample}, the jets in 3C~305 and IC~5063 are expected to strongly intersect the disk (i.e., small jet--disk angles), whereas in NGC~1052 the intersection is minimal (jet--disk angle of $\sim 10^\circ$). The remaining galaxies exhibit intermediate configurations. However, the objects in the sample are early-type galaxies, where the presence of a well-ordered disk structure is uncertain.

Another possible interpretation is that such wide-angle outflows are inherently common in AGNs but remain difficult to detect unless the data reach very high sensitivity, as the emission from shocked gas is easily overwhelmed by the much stronger photoionized component within the AGN ionization structure. This scenario is in agreement with theoretical models in which radiation pressure drives outflows with nearly spherical geometries for low black hole spins \citep{Ishibashi15,Ishibashi19}, and is reminiscent of the wide-angle outflows observed in luminous AGNs \citep[e.g.,][]{zakamska14,Harrison14,kakkad20}.

Considering that all galaxies in our sample host radio jets, the most plausible explanation for the enhanced gas turbulence observed perpendicular to the jets is that it is driven by the interaction of the jet as it propagates through the galactic disk. However, the small size of our sample and the uncertain presence and orientation of galactic disks in some objects preclude a more robust assessment of this scenario.

\subsection{Interaction of jets with the ISM: Shocked H$_2$ emission}\label{sec:disc-lr}

\textit{Spitzer} observations have revealed that RL AGNs exhibit H$_2$/PAH ratios up to two orders of magnitude higher than those in normal star-forming galaxies, indicating that the radio jet plays a dominant role in regulating the molecular gas excitation \citep[e.g.,][]{Ogle07,Ogle10,Nesvadba10}. In these systems, jet-ISM interactions and jet-driven outflows inject kinetic energy into the surrounding gas, producing shocks that enhance H$_2$ emission and can suppress star formation in the central regions \citep{Guillard12,Lanz16,Lanz15,Ogle24}. More generally, AGN host galaxies, including both RL and radio-quiet objects, exhibit elevated H$_2$/PAH ratios and higher H$_2$ excitation temperatures compared to star-forming galaxies \citep{Petric18,lambrides19}, with this ratio correlating with optical shock tracers such as [O I]$\lambda$6300 \citep{rogemar20_spitzer}. ALMA observations of cold molecular gas further show that highly disturbed gas is often co-spatial with strong H$_2$ emission and, when present, radio jets \citep{Nesvadba21,Ogle24}, confirming that shocks driven by outflows and jets inject energy into the ISM, enhancing the H$_2$ emission.

The use of the H$_2$/PAH\,11.3 ratio as a shock diagnostic is well established in the literature \citep[e.g.,][]{Ogle10,Guillard12,hill14,rogemar25_jwst}, with enhanced values commonly interpreted as evidence of shock-heated molecular gas relative to PAH emission associated with star formation. While PAH emission primarily traces the dust content and UV radiation field, shocks and turbulence produced by AGNs, particularly those hosting jets, can both enhance H$_2$ emission and partially destroy or suppress PAH molecules, modifying their physical and chemical properties \citep{Diamond-Stanic10,Appleton17,Zhang22}. High-energy radiation can ionize and destroy PAH molecules \citep{garcia-bernete22,Garcia-Bernete24}, while jet-driven shocks can induce fragmentation, erosion, and structural changes \citep{Micelotta10a,Micelotta10b}. However, variations in PAH strength unrelated to shocks, such as differences in the radiation field or dust properties, can introduce additional scatter in the H$_2$/PAH ratio. For this reason, we interpret this ratio in combination with other diagnostics, such as gas kinematics and emission-line ratios, rather than in isolation.

JWST observations further reinforce the role of shocks in driving H$_2$ mid-IR emission in nearby AGN host galaxies, which are mostly radio-quiet. In particular, \citet{rogemar26_CAB} analyzed MIRI/MRS observations of 54 nearby galaxies, including 41 AGN and 13 star-forming systems, comparing the observed line ratios and kinematics with shock and photoionization models. They found that shocks produced by radiation driven outflows or jet-cloud interactions are required to reproduce both the observed H$_2$ emission and the low-ionization emission line fluxes. Detailed studies of small AGN samples that directly mapped the spatial distribution of the H$_2$/PAH ratio consistently conclude that shocks play an important role in producing the observed warm H$_2$ emission, finding the highest H$_2$/PAH values in AGN-dominated regions and the lowest in star-forming regions \citep{U22,Garcia-Bernete24}, and reporting correlations between H$_2$/PAH, shock-tracing line ratios, and line width, which indicate that the H$_2$ emission in kinematically disturbed regions is primarily driven by shock-heated gas \citep{rogemar25_jwst}. \citet{Dasyra24} report high H$_2$/PAH ratios associated with a bow-shock region in IC\:5063 (one of the targets in our sample) and suggest that the enhanced H$_2$ emission may be driven by shocks or, alternatively, that the PAH emission may be suppressed due to PAH destruction by the jet. Furthermore, \citet{Zhang26} analyzed the nuclear H$_2$ emission in four low-luminosity AGNs using JWST/MIRI MRS and found that the molecular gas is not fully thermalized, indicating additional excitation from slow, likely jet-driven shocks. However, shocks are not universally required to reproduce the H$_2$ emission in AGNs; for instance, in NGC~5506 the observed H$_2$ emission is predominantly associated with shock excitation, whereas in NGC~3081 the H$_2$ emission is fully consistent with being driven by the radiation field of the central AGN \citep{Delaney25}. 

Here we investigate the spatial correlations between radio jets, the H$_2$/PAH ratio, and other diagnostic line ratios, providing new insight into the role of jets in driving the observed H$_2$ emission and elevated H$_2$/PAH values in RL AGNs. Results of the correlation tests are summarized in Fig.~\ref{fig:correlation}. We find clear correlations between the H$_2$/PAH ratio and the H$_2$ excitation temperature for all galaxies, both along the jet and perpendicular to it, except for NGC\:1052 along the jet. Among the galaxies where correlations are present, those measured along the jet are systematically stronger than those measured perpendicular to it. 

Correlations between the H$_2$ and [Fe\,{\sc ii}] emission are found along the jet for the same galaxies, whereas perpendicular to the jet no correlation is observed for 3C~305 and NGC\:1052. As higher excitation temperatures are expected in shocked H$_2$ gas \citep[e.g.,][]{Shull78,Hollenbach89,Flower10,Guillard09,Ogle10} and [Fe\,{\sc ii}] is a well-established tracer of shocks \citep[e.g.,][]{Oliva90,Oliva01,Forbes93,Hartigan04}, these trends indicate that shocks contribute to the excitation of the H$_2$ emission. A possible explanation for the lack of correlation perpendicular to the jet in 3C~305 is that the [Fe\,{\sc ii}] emission is spatially limited in this direction ($\sim1\arcsec$). In NGC~1052, both tracers show roughly circular distributions and no clear correlation (either along or perpendicular to the jet), possibly due to its low radio power and small jet--disk angle ($\sim10^\circ$; Table~\ref{tab:sample}), indicating weak jet--ISM interaction. In addition, H$_2$ and [Fe\,{\sc ii}] are not expected to arise from the same shocks, as the fast shocks required to produce [Fe\,{\sc ii}] emission would dissociate H$_2$ molecules, with the H$_2$ likely tracing post-shock regions \citep{Hollenbach89,Mouri00,rogemar26_CAB}. Finally, we note that, for Cygnus~A, a much stronger correlation between H$_2$ and [Fe\,{\sc ii}] is observed perpendicular to the jet, which can be explained if part of the [Fe\,{\sc ii}] emission along the jet is due to AGN photoionization \citep{Ogle25}.

The [Ne\,{\sc iii}]/[Ne\,{\sc ii}] ratio is primarily a tracer of the AGN radiation field \citep[e.g.,][]{Melendez08,Weaver10}, although these lines can also be excited by shocks, typically producing much fainter emission than that generated by AGN photoionization \citep{Feltre23,HermosaMunoz25}.
 We find anticorrelations between the H$_2$/PAH ratio and the [Ne\,{\sc iii}]/[Ne\,{\sc ii}] ratio along the jet direction for 3C\:305, Cygnus\:A, and IC\:5063, whereas no correlation is observed for 3C\:293 and NGC\:1052.  For 3C~305, the anticorrelation is weak, with the lowest H$_2$/PAH values roughly surrounding regions of higher [Ne\,{\sc iii}]/[Ne\,{\sc ii}] (Fig.~\ref{fig:maps_3C305}). In IC~5063, the highest H$_2$/PAH values are spatially associated with the radio knots \citep{Dasyra24}, supporting a scenario in which jet-driven shocks enhance the H$_2$ emission, while the decrease in H$_2$ in regions of high [Ne\,{\sc iii}]/[Ne\,{\sc ii}] (Fig.~\ref{fig:maps_3C305}) suggests partial suppression by the AGN radiation field. In Cygnus~A, the PAH emission is compact and only slightly elongated perpendicular to the jet (Fig.~\ref{fig:maps_CygnusA}), making the interpretation of observed anticorrelation more challenging.  This suggests that the enhancement of H$_2$/PAH along the jet cannot be explained solely by PAH destruction; rather, an intrinsic increase in the H$_2$ emission is also required to account for these correlations. These results also indicate that the shocks responsible for the H$_2$ excitation are associated with jet–cloud interactions or jet-driven outflows, rather than radiation-driven outflows. The only exception is Centaurus~A, where we observe a correlation between H$_2$/PAH and [Ne\,{\sc iii}]/[Ne\,{\sc ii}] along the jet direction, although with a smaller correlation coefficient than that found for H$_2$/PAH versus H$_2$ excitation temperature and [Fe\,{\sc ii}]/PAH. The enhancement of $T_{\rm H_2(3,1)}$ toward the northeast, co-spatial with the near side of the jet \citep{Hardcastle03,Neumayer07}, indicates that the jet is shaping the colder H$_2$ gas distribution and that shocks associated with the radio jet are the main excitation mechanism of the H$_2$, as discussed in detail by \citet{Evangelista26}.

Perpendicular to the jet, correlations between the H$_2$/PAH ratio and [Ne\,{\sc iii}]/[Ne\,{\sc ii}] are observed in three galaxies (Centaurus A, Cygnus A, and NGC 1052), while no correlation is found in the other three. A possible interpretation is that, in these three galaxies, part of the ionized gas emission also arises from shock-related processes. Since photoionized gas produces [Ne\,{\sc iii}] emission more efficiently than shock-ionized gas, any contribution from shocks to the [Ne\,{\sc iii}] lines within the AGN ionization region is likely masked by the dominant photoionization, in a manner similar to what has been observed for [O\,{\sc iii}] \citep{rogemar21_te}, as [Ne\,{\sc iii}] and [O\,{\sc iii}] have comparable ionization potentials (41.0 and 35.1 eV, respectively). Another possibility is a contribution from X-ray irradiation to the H$_2$ emission, since X-rays can escape through the AGN torus and heat the molecular gas in extended regions, potentially enhancing the H$_2$ emission independently of shocks \citep{Dors12,rogemar_n5929}.

We also explored the correlations for the same parameters along and perpendicular to the disk plane, finding no systematic trends with disk orientation, suggesting that any effect related to the disk orientation is secondary compared to that associated with the jet.
 In summary, we find that the H$_2$/PAH ratio is strongly correlated with H$_2$ excitation temperature and [Fe\,{\sc ii}] emission, particularly along the radio jet, indicating that shocks play a major role in powering the observed H$_2$ emission. The combination of these trends with the behavior of the [Ne\,{\sc iii}]/[Ne\,{\sc ii}] ratio supports a scenario in which jet-driven shocks are the primary driver of the enhanced H$_2$/PAH values in RL AGNs \citep{Ogle07,Ogle10,rogemar25_jwst}.

\section{Conclusions} \label{sec:conc}

We analyzed JWST/MIRI MRS observations of seven nearby RL AGN host galaxies to investigate the impact of radio jets on gas turbulence and H$_2$ excitation. 
 Our main conclusions are:

\begin{itemize}
\item Six out of the seven galaxies in our sample exhibit enhanced gas velocity dispersion in regions perpendicular to the radio jet axis.\ This is  observed in both molecular (3C\:305, IC\:5063, NGC\:1052, and M\:87) and ionized (3C\:293, 3C\:305, Centaurus~A, and IC\:5063) gas and 
 indicates that jet--ISM interactions are not confined to the collimated jet channel but also drive turbulence on wider angular scales. The only exception is Cygnus A, which has the highest radio luminosity of the sources in the sample. The enhanced turbulence is interpreted as being associated with radio jets launched at small angles relative to the disk plane, which lose momentum through interactions with dense clouds and drive wide-angle outflows that emerge approximately perpendicular to the disk.

\item Among the seven galaxies, only M87 does not show extended emission in the PAH\:11.3\:$\mu$m feature.  Strong correlations are found between the H$_2$\,S(3)/PAH\,11.3$\:\mu$m ratio and the H$_2$ excitation temperature in most galaxies, both along and perpendicular to the jet direction, with systematically stronger correlations along the jet. This indicates that shocks play a major role in the observed H$_2$ emission. The only exception is NGC\:1052, which hosts the lowest-power jet in the sample.

\item Additional correlations between the H$_2$ and [Fe\,{\sc ii}] emission along the jet further support a shock-driven excitation of the molecular gas, as [Fe\,{\sc ii}] is a well-established tracer of shocks in dense neutral gas. NGC\:1052 is again an exception; AGN radiation is likely the dominant mechanism driving its H$_2$ excitation.

\item Anticorrelations between the H$_2$/PAH ratio and the [Ne\,{\sc iii}]/[Ne\,{\sc ii}] ratio observed along the jet in several galaxies indicate that the enhancement of H$_2$/PAH cannot be explained solely by PAH destruction. Instead, an intrinsic increase in the H$_2$ emission due to jet-driven shocks is required. In regions perpendicular to the jet, the shocks are likely associated with the lateral expansion of jet-driven outflows, as indicated by the higher gas turbulence perpendicular to the jet observed in six out of the seven sources.
\end{itemize}

Overall, our results demonstrate that radio jets play a central role in shaping both the kinematics and excitation of the multiphase ISM on scales of hundreds of parsecs to a few kiloparsecs, driving shocks that enhance turbulence and molecular hydrogen emission in the nuclear region of RL AGNs. Future studies will help elucidate this picture, particularly through improved constraints on the jet orientation relative to the disk plane, which will enable us to better characterize the jet--disk coupling, as well as through analyses of larger samples to explore how the impact of the jet on the ISM relates to radio power and other global AGN properties.

\begin{acknowledgements}
The authors thank the anonymous referee for the careful reading of the manuscript and for the constructive comments, which helped improve the clarity and interpretation of the results. This work is based on observations made with the NASA/ESA/CSA \textit{James Webb} Space Telescope. The data were obtained from the Mikulski Archive for Space Telescopes (MAST) at the Space Telescope Science Institute, which is operated by the Association of Universities for Research in Astronomy, Inc., under NASA contract NAS 5-03127 for JWST.  The AI tool ChatGPT (GPT-4.5) was used to assist in debugging the code used to produce the figures and to refine the wording of some sentences in the manuscript. This research made use of Astropy, a community-developed core Python package for Astronomy \citep{Astropy13,Astropy22}. 

RAR  acknowledges the support from the Conselho Nacional de Desenvolvimento Científico e Tecnológico (CNPq; Projects 303450/2022-3, and 403398/2023-1), the Coordenação de Aperfeiçoamento de Pessoal de Nível Superior (CAPES; Project 88887.894973/2023-00), and Fundação de Amparo à Pesquisa do Estado do Rio Grande do Sul (FAPERGS; Project 25/2551-0002765-9). 
LC acknowledges support by grant PIB2021-127718NB-I00 from the Spanish
Ministry of Science and Innovation/State Agency of Research 
MCIN/AEI/10.13039/501100011033 and by “ERDF A way of making Europe”.

AAH and LHM acknowledge financial support by the grant PID2021-124665NB-I00 funded by the Spanish Ministry of Science and Innovation and the State Agency of Research MCIN/AEI/10.13039/501100011033 PID2021-124665NB-I00 and ERDF A way of making Europe. 

GLSO, SAS, MSZM, and LRV acknowledge financial support from Coordena\c c\~ao de Aperfei\c coamento de Pessoal de N\'ivel Superior (CAPES; Finance Code 001).

EB acknowledges support from the Spanish grants PID2022-138621NB-I00 and PID2021-123417OB-I00, funded by MCIN/AEI/10.13039/501100011033/FEDER, EU. 

MB acknowledges support from the Juan de La Cierva scholarship with reference JDC2023-052684-I, funded by MICIU/AEI/10.13039/501100011033 and FSE+.

RR acknowledges support from CNPq (Proj. 445231/2024-6,311223/2020-6, 404238/2021-1, and 310413/2025-7), FAPERGS (Proj. 19/1750-2 and 24/2551-0001282-6) and  CAPES (88881.109987/2025-01).

NLZ acknowledges support from  STScI through program JWST-GO-01928.

GSC acknowledges support from DFG under grant CO 3160/1-1.

AA is co-funded by the European Union (Widening Participation, ExGal-Twin, GA 101158446). 

CRA, AA and MB acknowledge support from the Agencia Estatal de Investigaci\'on of the Ministerio de Ciencia, Innovaci\'on y Universidades (MCIU/AEI) under the grant ``Tracking active galactic nuclei feedback from parsec to kiloparsec scales'', with reference PID2022$-$141105NB$-$I00 and the European Regional Development Fund (ERDF). 

CR acknowledges support from SNSF Consolidator grant F01$-$13252, Fondecyt Regular grant 1230345, ANID BASAL project FB210003 and the China-Chile joint research fund.

IGB is supported by the Programa Atracci\'on de Talento Investigador ``C\'esar Nombela'' via grant 2023-T1/TEC-29030 funded by the Community of Madrid, and acknowledges support from the research project PID2024-159902NA-I00 funded by the Spanish Ministry of Science and Innovation/State Agency of Research (MCIN/AEI/10.13039/501100011033) and FSE+

SGB acknowledges support from the Spanish grant PID2022-138560NB-I00, funded by MCIN/AEI/10.13039/501100011033/FEDER, EU.

MPS acknowledges support from grants RYC2021-033094-I, CNS2023-145506, and PID2023-146667NB-I00 funded by MCIN/AEI/10.13039/501100011033 and the European Union NextGenerationEU/PRTR.

\end{acknowledgements}

\bibliography{aa59627-26_final}

\begin{appendix}

\section{Two-dimensional maps} \label{appendix:2D}

\begin{figure*}
   \centering
   \includegraphics[width=0.95\textwidth]{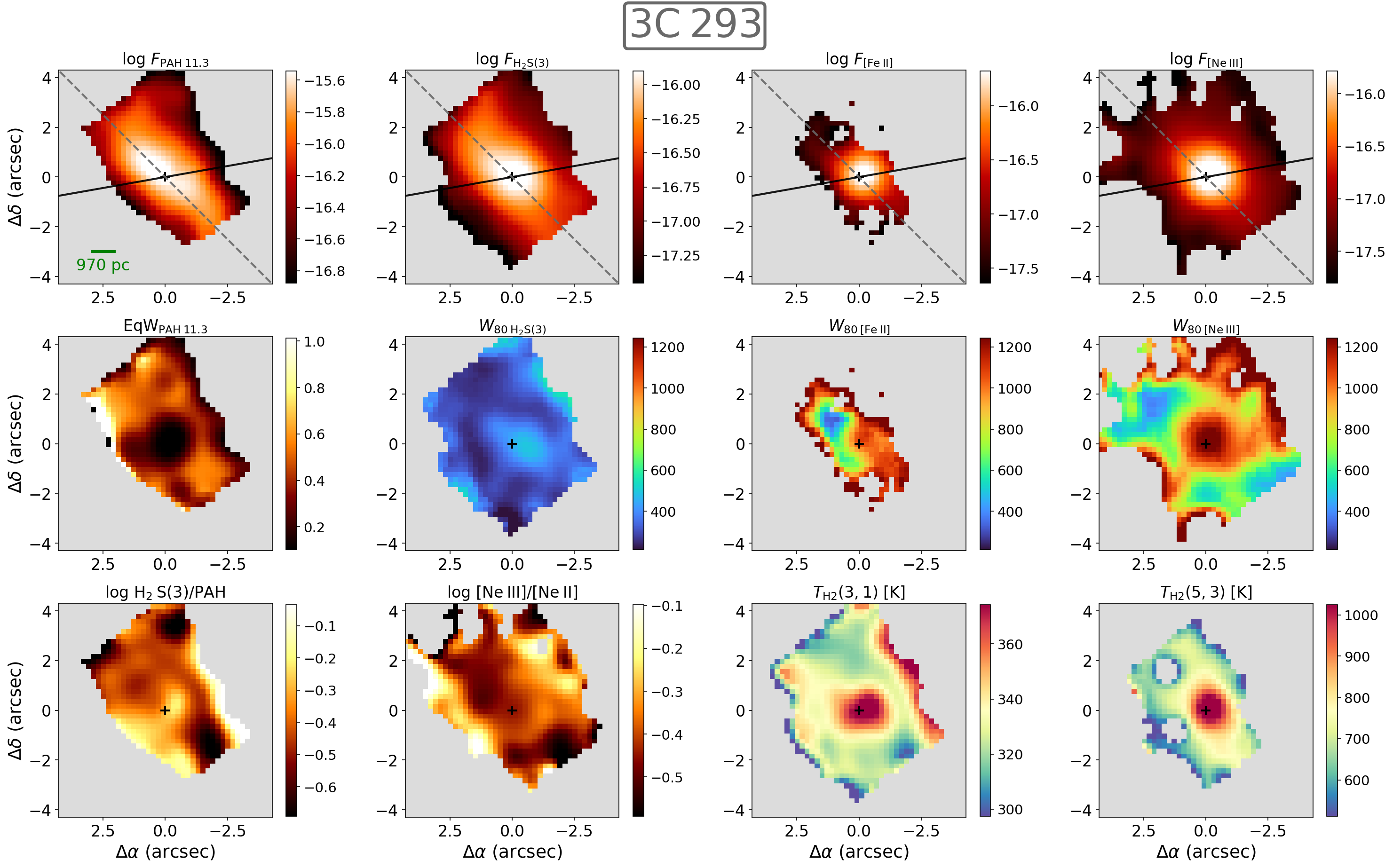}   
  
      \caption{Maps for 3C\:293. \textit{Top row}: Flux distributions of the PAH\,11.3$\:\mu$m feature, H$_2$ S(3), [Fe\,{\sc ii}]\,5.34$\:\mu$m, and [Ne\,{\sc iii}]\,15.56$\:\mu$m emission lines. The color bars indicate the flux values on a logarithmic scale, in units of erg\,s$^{-1}$\,cm$^{-2}$. The solid line shows the radio jet orientation, and the dashed line indicates the disk orientation (Table~\ref{tab:sample}). \textit{Middle row} (left to right): PAH\,11.3$\:\mu$m EW map, with the color bar indicating values in $\mu$m, and the H$_2$ S(3), [Fe\,{\sc ii}], and [Ne\,{\sc iii}] $W_{\rm 80}$ maps, in units of km\,s$^{-1}$. \textit{Bottom row }(left to right): log\,H$_2$\,S(3)/PAH\,11.3$\:\mu$m and log\,[Ne\,{\sc iii}]\,15.56$\:\mu$m/[Ne\,{\sc ii}]\,12.81$\:\mu$m flux ratio maps, as well as the H$_2$ excitation temperature maps derived from the H$_2$ S(3) and S(1), and S(3) and S(5) emission lines. The crosses mark the position of the galaxy nucleus, while the gray regions correspond to spaxels where the emission lines are not detected at A/N $> 5$ (or A/N $> 3$ for the PAH feature), or are not covered by the MIRI FoV.
 }
         \label{fig:maps_3C293}
   \end{figure*}

\begin{figure*}
   \centering
   \includegraphics[width=0.95\textwidth]{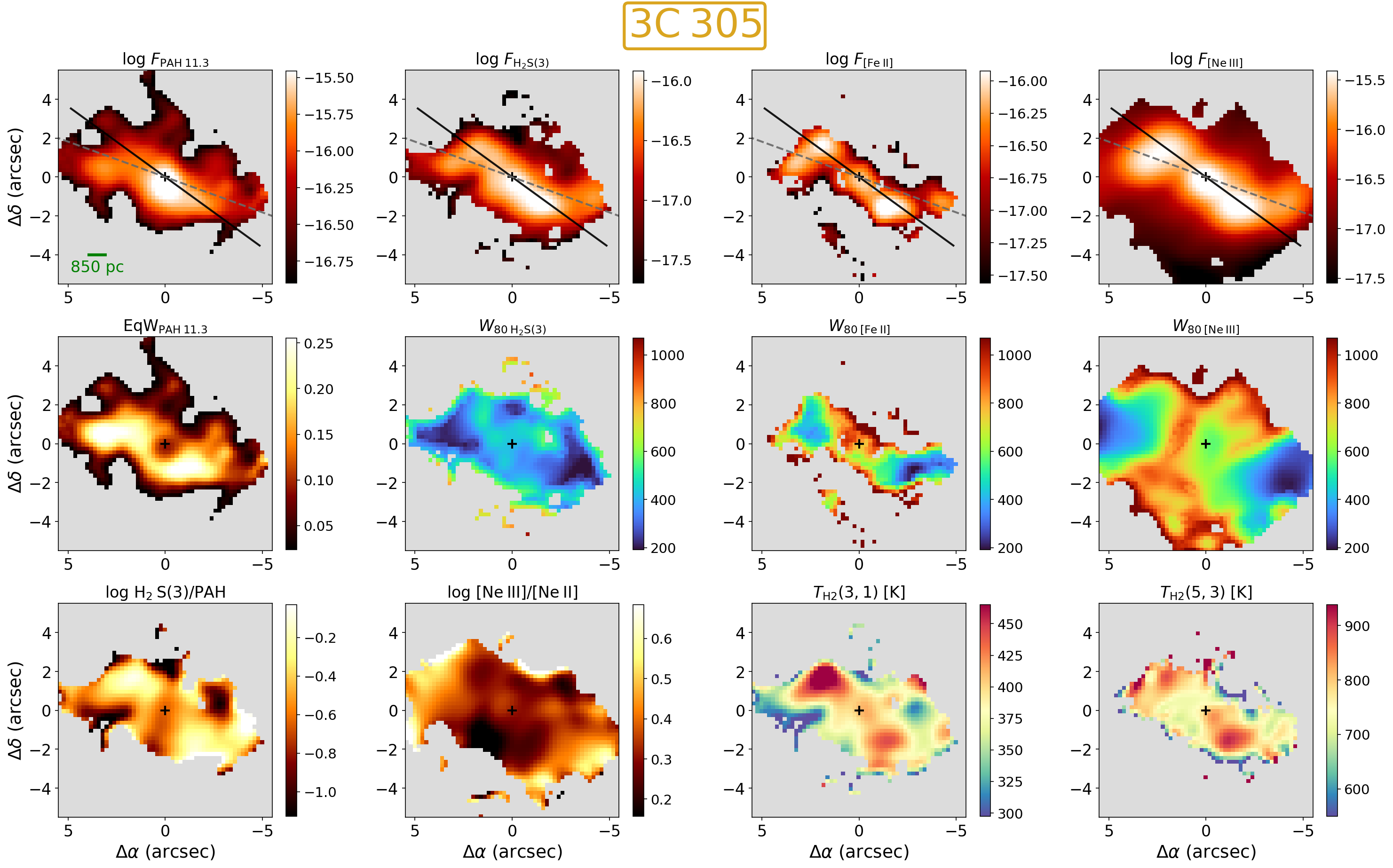}   
  
      \caption{Same as Fig.~\ref{fig:maps_3C293} but for 3C\:305.}

         \label{fig:maps_3C305}
   \end{figure*}

\begin{figure*}
   \centering
   \includegraphics[width=0.95\textwidth]{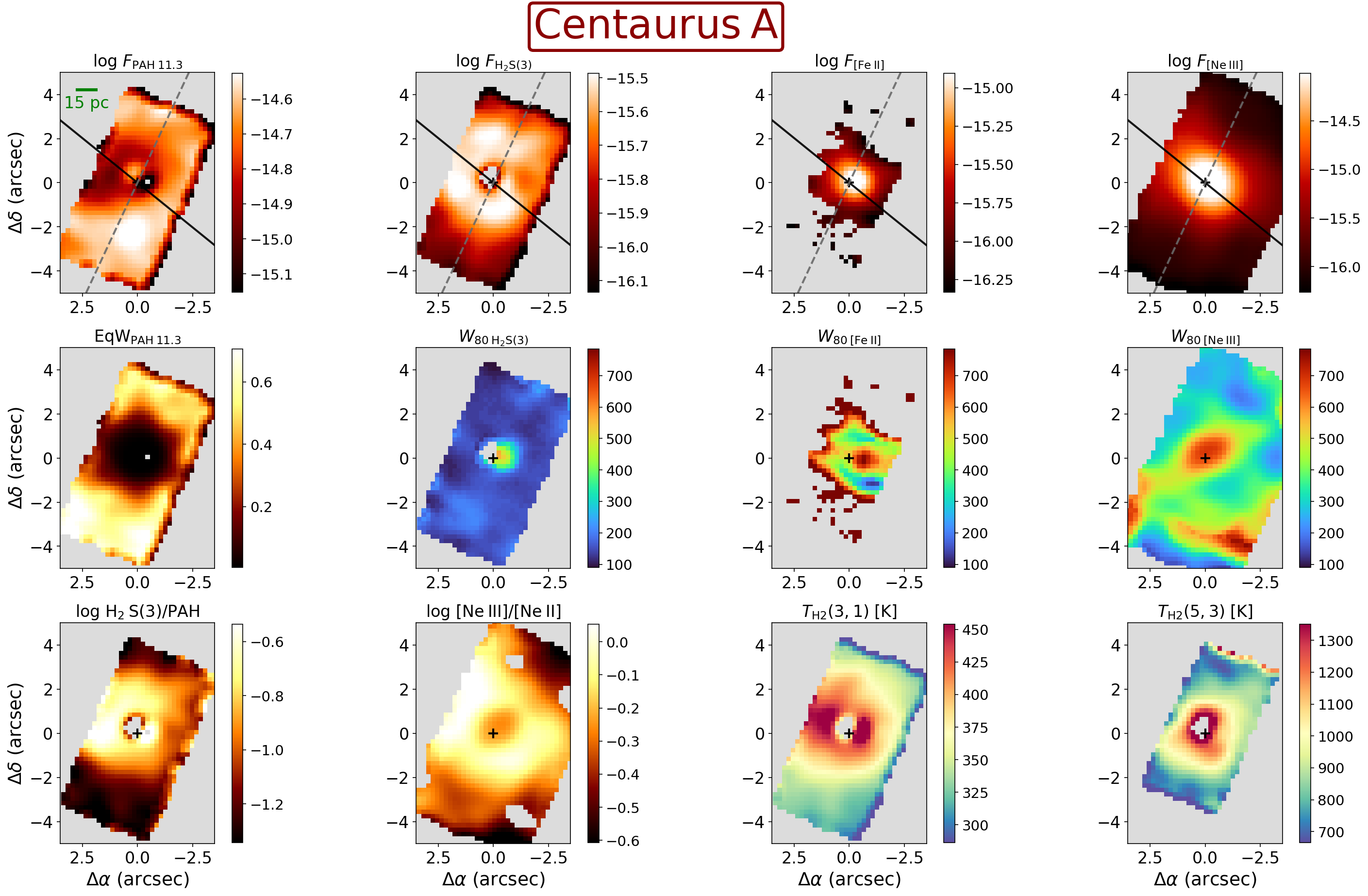} 
      \caption{Same as Fig.~\ref{fig:maps_3C293} but for Centaurus A.}
         \label{fig:maps_CentaurusA}
   \end{figure*}

\begin{figure*}
   \centering
   \includegraphics[width=0.95\textwidth]{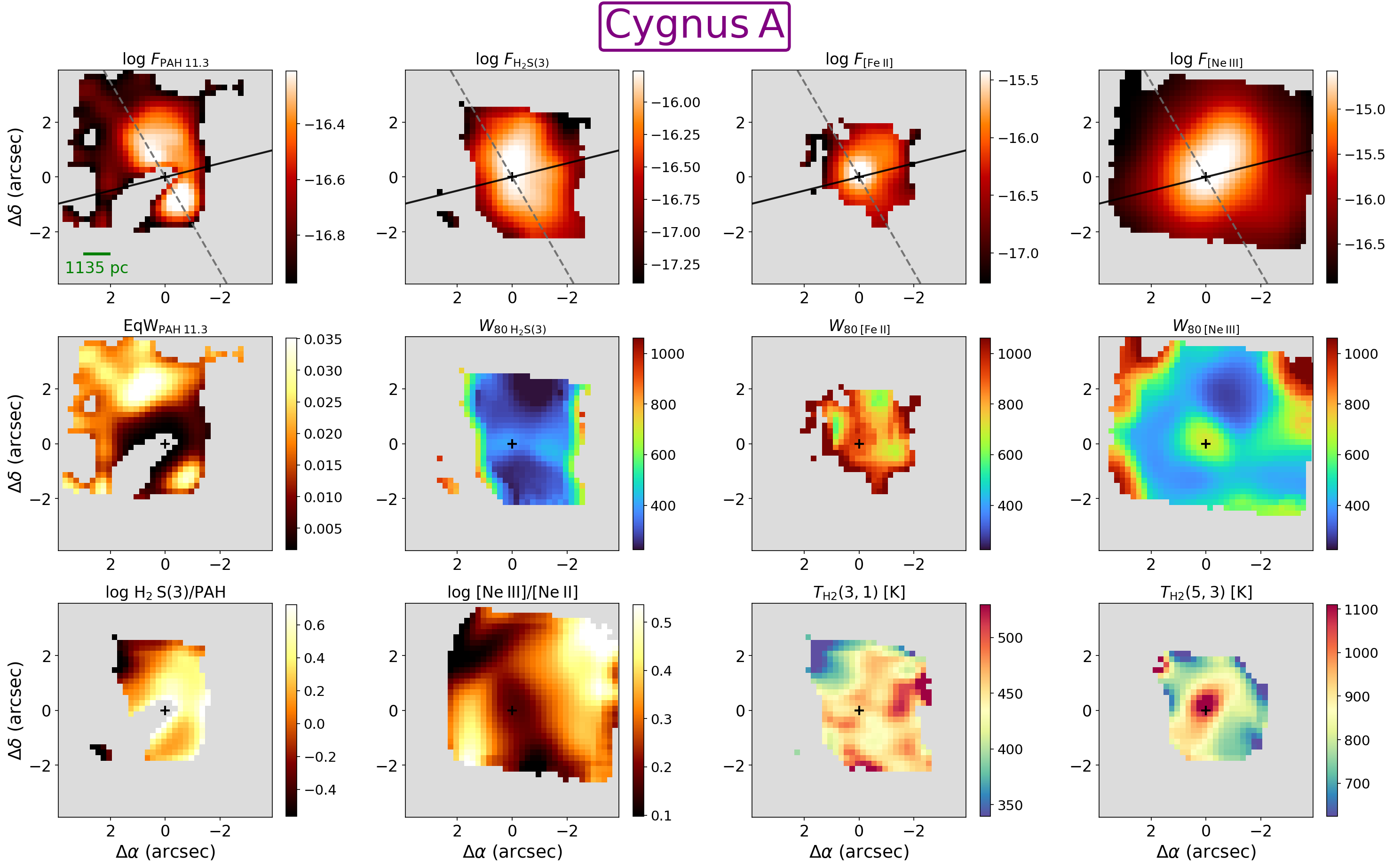}

      \caption{Same as Fig.~\ref{fig:maps_3C293} but for Cygnus A.}
         \label{fig:maps_CygnusA}
   \end{figure*}

\begin{figure*}
   \centering
   \includegraphics[width=0.95\textwidth]{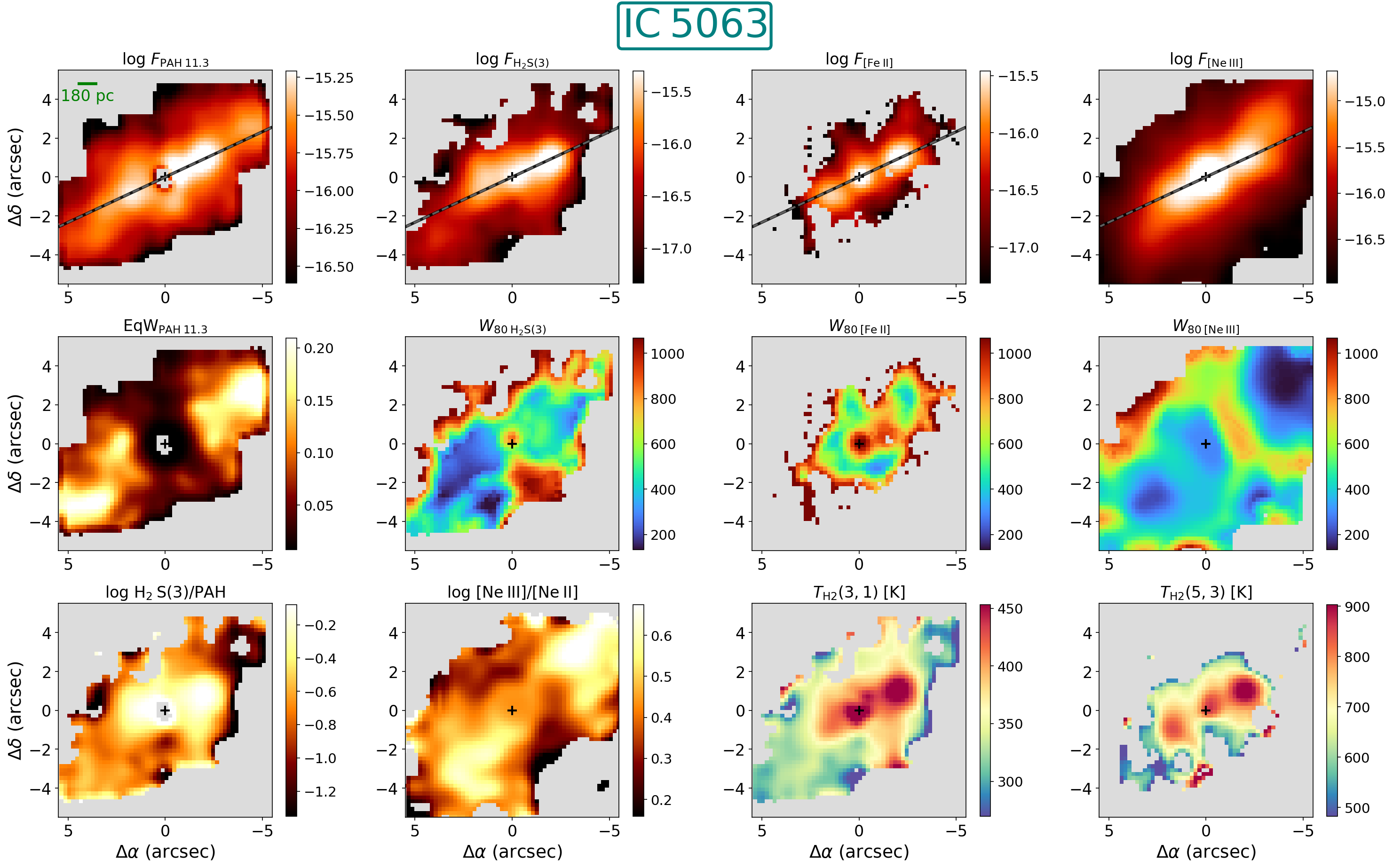}
      
      \caption{Same as Fig.~\ref{fig:maps_3C293} but for IC\:5063.}
         \label{fig:maps_IC5063}
   \end{figure*}

\begin{figure*}
   \centering
   \includegraphics[width=0.95\textwidth]{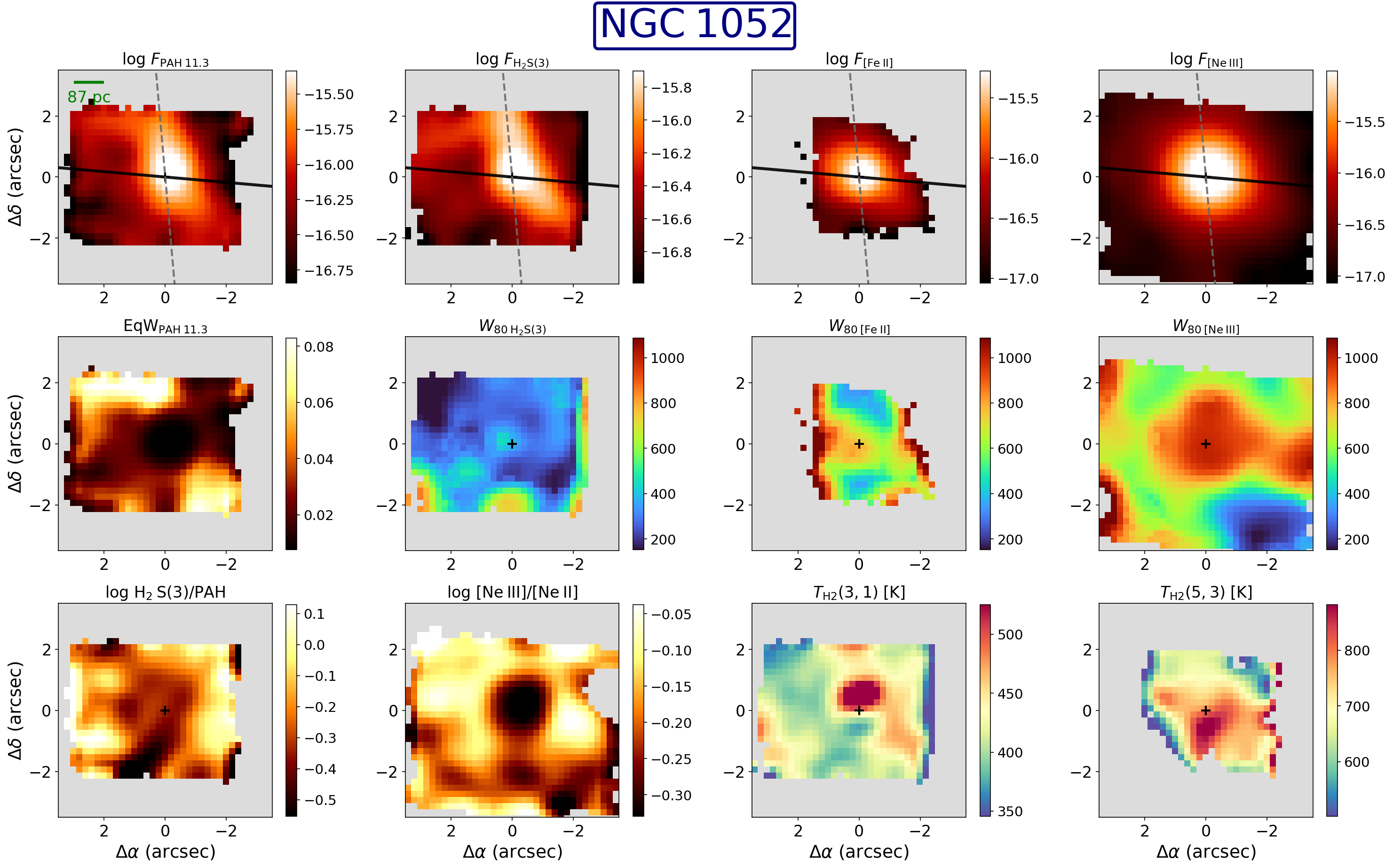}       
      \caption{Same as Fig.~\ref{fig:maps_3C293} but for NGC\:1052.}
         \label{fig:maps_NGC1052}
   \end{figure*}

\begin{figure*}
   \centering
   \includegraphics[width=0.73\textwidth]{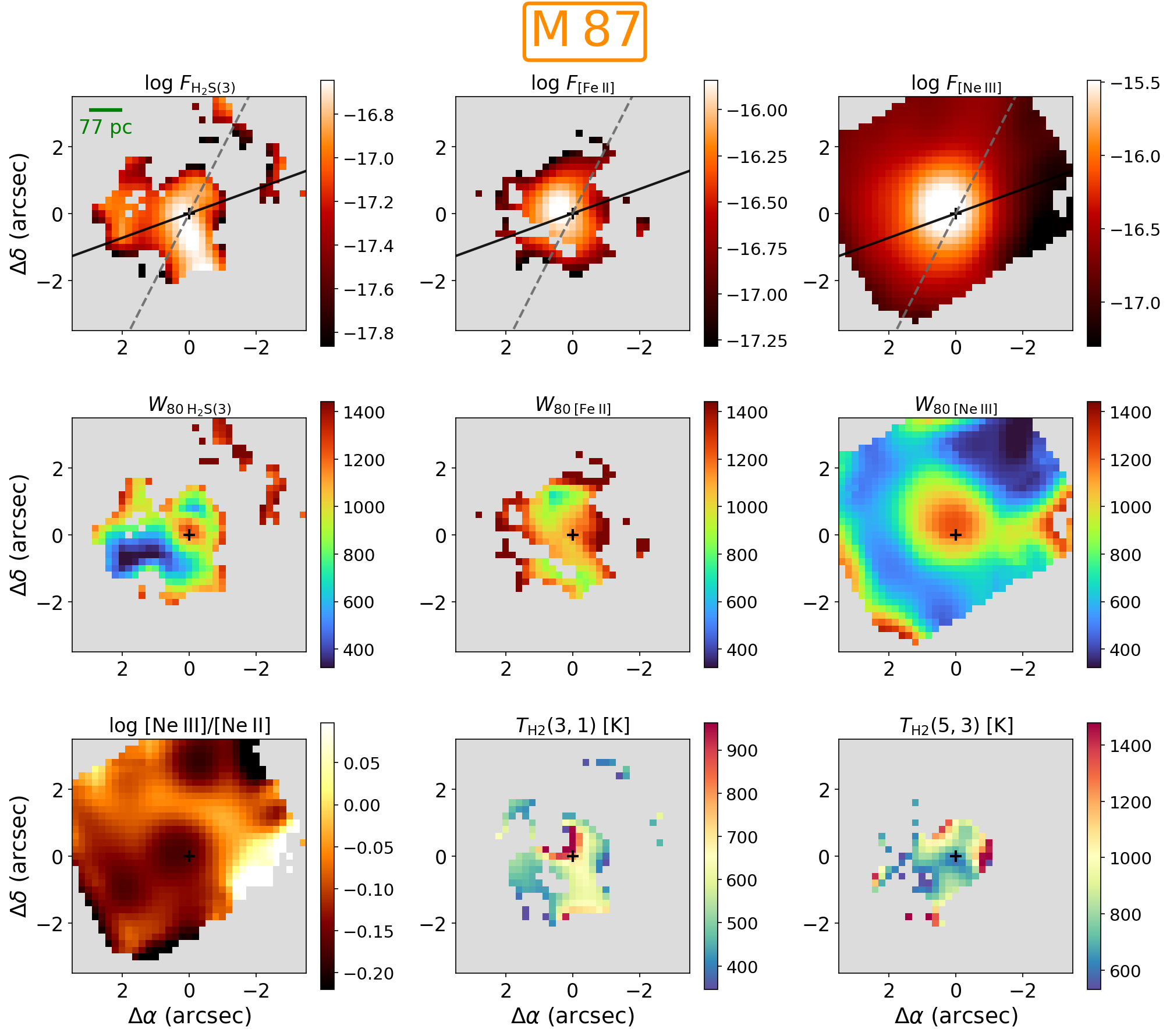}      
      \caption{Same as Fig.~\ref{fig:maps_3C293} but for M\:87. Maps that use the PAH\,11.3$\:\mu$m feature are not shown, as it is not detected in this galaxy.}
         \label{fig:maps_M87}
   \end{figure*}

 Figures~\ref{fig:maps_3C293}--\ref{fig:maps_M87} show the two-dimensional maps for the galaxies 3C\:293, 3C\:305, Centaurus~A, Cygnus~A, IC~5063, NGC~1052 and M~87, respectively. Below, we describe these maps and place them in the context of previous radio and JWST-based studies of the galaxies in our sample.

\noindent{\bf 3C\:293} (Fig.~\ref{fig:maps_3C293}): 3C\:293 is a restarted radio galaxy, exhibiting a double-double morphology with both inner and outer lobes, and classified as a Fanaroff-Riley I \citep[FR~I;][]{FR74}  type \citep{Chiaberge99,Beswick04,Floyd06,Kukreti22}. The large-scale radio emission extends up to approximately 200~kpc, oriented approximately along the northwest-southeast direction, whereas the inner lobes, observed on a $\sim$4~kpc scale, are oriented along a position angle (PA) of $\approx$100$^\circ$, offset from the large-scale lobe orientation by $\sim$35$^\circ$ \citep{Bridle81,Machalski16,Kukreti22}. The radio continuum morphology indicates that the jet is bent, likely by interactions with the dense structures in the ISM  \citep{Beswick04}. 

The first spatially resolved maps of H$_2$ emission in 3C~293 were presented by \citet{Lanz15} using \emph{Spitzer} observations. By combining these data with deep \emph{Chandra} observations, they found comparable masses of X-ray–emitting hot gas and warm molecular gas, concluding that both originate from a shock-driven, multiphase ISM.  The interaction between the jet and the ISM drives multiphase outflows, observed in neutral gas \citep{Morganti03, Mahony13}, in the ionized phase \citep{Emonts05, Mahony16,rogemar23_ugc}, in the hot molecular component traced by near-IR H$_2$ emission \citep[e.g.,][]{deMellos25}, and in the warm molecular gas phase revealed by mid-IR rotational H$_2$ lines \citep{henrique24}. Previous results based on MIRI/MRS observations of 3C\,293 show that the warm molecular outflow rate exceeds that derived from the optical ionized gas by about one order of magnitude \citep{henrique24}, a difference that can be explained by the fact that the mid-IR data probe a much dustier component of the outflow, consistent with the high extinction observed in this galaxy \citep{deMellos25}. This high extinction also accounts for the $\sim 0.6\arcsec$ offset between the peak emission in the optical and in the IR \citep{deMellos25}.  These observations further reveal that shocks driven by the radio jet or by the outflow itself dominate the excitation of the H$_2$ emission \citep{rogemar25_jwst}. 

As seen in Fig.~\ref{fig:maps_3C293}, the PAH\,11.3\,$\mu$m and H$_2$\,S(3) emission are elongated along the galaxy’s major-axis orientation \citep[PA = $50^\circ$;][]{2mass}, showing a westward bend on the northern side, near the edges of the field of view (FoV). The [Fe\,{\sc ii}] emission displays a round morphology at the highest intensity levels, with fainter emission extending along the disk, while the [Ne\,{\sc iii}] emission is detected across the entire FoV, showing an overall rounder distribution but slightly elongated along the disk. The PAH EW map shows the lowest values at the nucleus, increasing outward, primarily along the disk. The H$_2$ $W_{\rm 80}$ map exhibits overall lower values than those observed in the ionized gas, with the highest values found at the nucleus and along the northwest–southeast direction, coinciding with the orientation of the large-scale radio jet \citep{Kukreti22}. A similar increase in velocity dispersion is also observed in ionized gas, reaching values up to 1200 km\,s$^{-1}$, while the lowest $W_{\rm 80}$ values are found to the northeast and southwest, along the disk orientation. The H$_2$\,S(3)/PAH\,11.3\,$\mu$m and [Ne\,{\sc iii}]\,15.56\,$\mu$m/[Ne\,{\sc ii}]\,12.81\,$\mu$m flux-ratio maps show the highest values along the northwest–southeast direction, while the lowest values are observed in the perpendicular direction. The H$_2$ excitation temperature maps display the highest temperatures at the nucleus, with the lowest values occurring along the disk orientation.

\noindent{\bf 3C\:305} (Fig.~\ref{fig:maps_3C305}): The radio continuum emission of 3C\,305 on kiloparsec scales reveals two symmetric jets aligned along a PA of 54$^\circ$, forming well-defined radio lobes separated by about 3.6~arcsec. Two low-brightness arms extend roughly perpendicular to the radio axis, emerging from the outer regions of the lobes. The total extent of the jet structure is only about 5~arcsec ($\sim$4~kpc), consistent with a small-scale FR~II morphology \citep{Heckman82,Jackson03,Morganti05,Hardcastle12}. 
\citet{Sebastian26} used JWST near- and mid-IR integral field spectroscopy and imaging to investigate the impact of the radio jet on the ISM of 3C~305. By combining molecular, ionic, and dust tracers, including rotational H$_2$ lines, PAH emission, and [Fe\,{\sc ii}], they find compelling evidence of shock excitation, particularly at the jet termination regions.  

The emission-line flux distributions in this galaxy show extended emission along the jet orientation, exhibiting an "S-shaped" morphology. While the ionized gas displays three distinct knots of emission, the H$_2$ emission is smoother and slightly stronger to the southwest of the nucleus. The PAH emission is also extended, peaking at the nucleus and appearing less collimated than the structures observed in the ionized and molecular gas maps. The PAH EW map shows the lowest values at the nucleus and in the northwestern side of the galaxy, while the highest values are seen to the southeast. The H$_2$ shows overall lower $W_{\rm 80}$ values compared to those of the ionized gas.  The lowest $W_{\rm 80}$ values ($<300~{\rm km~s^{-1}}$) for all emission lines are observed along the radio-jet orientation, while the highest ones ($\sim600~{\rm km~s^{-1}}$ for H$_2$ and $\sim900$--$1000~{\rm km~s^{-1}}$ for the ionized gas) are seen in the direction perpendicular to the jet. Both line ratios, H$_2$\,S(3)/PAH\,11.3$\:\mu$m and [Ne\,{\sc iii}]\,15.56$\:\mu$m/[Ne\,{\sc ii}]\,12.81$\:\mu$m, show increasing values with distance from the nucleus along the jet direction, while lower values are observed perpendicular to the jet orientation. Similarly, higher H$_2$ excitation temperatures are found along the jet direction.

\noindent{\bf Centaurus\:A} (Fig.~\ref{fig:maps_CentaurusA}): Centaurus\:A is the nearest radio galaxy and is widely regarded as the prototype of FR\:I radio galaxies \citep{Israel98}. On parsec scales, well‑collimated radio jets emerge from the nucleus, maintaining their collimation for several kiloparsecs before expanding into plumes \citep{Clarke92,Israel98,Hardcastle03}. On larger scales, Centaurus~A exhibits giant lobes extending up to $\sim 500$~kpc. These lobes are oriented predominantly north--south but show complex structures, including changes in PA, filamentary features, and a pronounced north--south asymmetry \citep{Morganti10,Feain11}.  

\citet{Alonso-Herrero25} used JWST MIRI/MRS to investigate in detail the multiphase gas distribution and kinematics of Centaurus\,A.  These authors report that the H$_2$ and ionized gas exhibit distinct flux distributions and kinematics, with the H$_2$ primarily associated with the disk and the ionized gas more extended along the radio jet. The ionized gas shows evidence of jet-driven outflows and bubbles, as well as shock excitation in the nuclear region.
\citet{Evangelista26} present a detailed study of the H$_2$ emission in Centaurus~A. They find that the low-excitation lines, in particular the H$_2$ 0--0 S(1) line, show a deficit of emission toward the northeast, aligned with the near side of the jet \citep{Hardcastle03}, indicating jet-driven shaping of the molecular gas. In contrast, the higher-excitation H$_2$ lines display a ring-like morphology with filamentary structures, similar to what was previously reported by \citet{Alonso-Herrero25}. In addition, \citet{Evangelista26}  fit the H$_2$ excitation diagrams on a spaxel-by-spaxel basis using a two-temperature model, producing maps of the warm and hot molecular gas temperatures ($T_{\rm warm}$ and $T_{\rm hot}$). They find that $T_{\rm warm}$ is enhanced in the region where the H$_2$ 0--0 S(1) emission decreases toward the northwest, while $T_{\rm hot}$ decreases radially from the nucleus outward. Based on these results, they conclude that shocks are the dominant excitation mechanism of H$_2$ in the circumnuclear disk of Centaurus~A.

The H$_2$ S(3) emission shows a ring-like structure within a radius of 2~arcsec, 
similar to the ring reported by \citet{Alonso-Herrero25} and \citet{Evangelista26} using the H$_2$ S(5) flux map, and attributed to emission from the nuclear disk located inside the cold molecular gas ring 
with a $9^{\prime\prime} \times 6^{\prime\prime}$ diameter \citep{Espada17}. The [Fe\,{\sc ii}] emission is detected only within the central $\sim2$~arcsec radius, while the [Ne\,{\sc iii}] flux distribution exhibits an ionization-cone morphology oriented along the radio axis, with lower flux values observed in regions perpendicular to the cone. The highest-intensity PAH emission appears to encase the H$_2$ S(3) emission, displaying a ring-like morphology with a radius consistent with that seen in the cold molecular gas \citep{Espada17} and in the lower-excitation H$_2$ S(1) line \citep{Alonso-Herrero25}.  The PAH EW show the highest values in the ring, while lower values are observed at smaller radii. The H$_2$ $W_{\rm 80}$ map shows overall low values ($<200~{\rm km~s^{-1}}$), 
while higher values are observed in the ionized gas, likely associated with a jet-driven bubble 
and outflow interacting with the galaxy's ISM, as reported by \citet{Alonso-Herrero25}. The H$_2$/PAH and [Ne\:{\sc iii}]/[Ne\:{\sc ii}] line ratios show the highest values within the ionization cone. In addition, the former exhibits some enhanced values to the northwest of the nucleus, in a region where the lowest [Ne\:{\sc iii}]/[Ne\:{\sc ii}] ratios are observed. Finally, the H$_2$ excitation temperature maps show decreasing values from the nucleus outward, but with a temperature enhancement to the northwest seen only in the low-excitation map, similarly to what is reported in \citet{Evangelista26}.

\noindent{\bf Cygnus\:A} (Fig.~\ref{fig:maps_CygnusA}): Cygnus\,A exhibits a prominent radio jet extending toward the northwest and a counter-jet toward the southeast of its nucleus. These jets are traced from subparsec scales out to distances of roughly 70~kpc, terminating in well-defined, edge-brightened lobes with prominent hotspots, characteristic of a classical FR~II radio galaxy \citep{Perley84,Linfield85,Carilli96}.  MIRI MRS and NIRSpec Integral Field Unit (IFU) data of Cygnus~A have been used by \citet{Ogle25} to study the multiphase gas dynamics of this galaxy. These authors show that the bipolar narrow-line region of Cygnus~A is shaped by the interaction of the radio jet with the ISM, exhibiting evidence of rotation around the radio axis and the presence of high-velocity outflows in the form of ``bullets'' reaching up to 2000~km~s$^{-1}$.  Their conclusions are based on the analysis of the spatially resolved emission structure and kinematics from the near-IR data, and measurements at selected locations for the mid-IR lines, without presenting the full spatial distribution of the mid-IR emission and kinematics. 

The PAH \,11.3$\:\mu$m emission in Cygnus~A is  weak, mostly detected perpendicular to the radio jet and the AGN ionization cone, and exhibits very low EWs (EW $<0.035~\mu$m). The H$_2$ emission is detected in an elongated structure along the north-south direction, approximately perpendicular to the radio jet, while the ionized gas emission is tracing the AGN ionization cone \citep{Rogemar_cygA,Ogle25}. The H$_2$ $W_{\rm 80}$ map shows values lower than $400~{\rm km~s^{-1}}$, while the ionized gas exhibits higher velocity dispersion, likely associated with the biconical outflow component \citep{Ogle25}. The [Ne\:{\sc iii}]/[Ne\:{\sc ii}] ratio map clearly delineates the wide-opening, X-shaped narrow-line region.  Higher H$_2$/PAH ratios also appear to be present within the ionization cone, although the corresponding map is limited by the small number of spaxels with detected PAH and H$_2$ emission. The H$_2$ excitation temperature map derived from the higher-excitation S(3) and S(5) lines shows the highest values at the nucleus, followed by a systematic decrease with increasing distance. The $T_{\rm H_2}(3,1)$ map displays a slightly different distribution, with the lowest temperatures to the northeast of the nucleus and higher values toward the west.

\noindent{\bf IC\:5063} (Fig.~\ref{fig:maps_IC5063}): IC~5063 presents a triple radio structure oriented along PA~$\sim 115^\circ$, with a total extent of approximately 1~kpc \citep{Morganti98}. The extranuclear radio components appear to be connected to the radio core by a low-brightness bridge, and diffuse emission is observed surrounding these components \citep{Morganti07}. Faint radio emission at 1.4~GHz is observed perpendicular to the main jet, which is not necessarily related to the AGN and may instead arise from starburst-driven winds \citep{Morganti07}.  

A detailed analysis of the mid-IR emission line flux distributions and kinematics of IC\:5063 is done by \citet{Dasyra24}, using JWST MIRI/MRS. These authors report a very complex kinematic structure in this source, with more than ten discrete regions showing outflows, associated with the radio lobes, the nucleus, a bicone perpendicular to the jet \citep[first reported by ][]{Dasyra15}, and a bubble moving against the disk. The outflows and evidence of shocks in regions distant from the radio jet are interpreted as the relics of a past jet or a weak radio that interacted with the ISM. Finally, they also reported higher H$_2$ temperatures in regions co-spatial with radio knots, suggesting shock-driven excitation or by cosmic rays.

In IC\:5063, all the emission-line flux distributions are well aligned along the jet direction, with emission observed over most of the MRS FoV. Close to the borders of the FoV to the northwest, the flux distributions show a bend toward the north, while a similar bend is seen to the south near the southeastern edge. This structure seems to trace the strongest emission in the optical lines, which are extended up to distances of $\sim15$\:arcsec from the nucleus along the jet orientation, presenting an X-shaped morphology at large distances from the center \citep{Fonseca-Faria23}. The PAH\,11.3\,$\mu$m flux distribution is similar to that of H$_2$, but this feature is not detected in the nucleus at A/N $> 3$. This region, however, is strongly affected by point spread function residuals, as extensively discussed by \citet{Dasyra24}. The PAH EW map shows the highest values along the direction of the jet, while the lowest values are observed perpendicular to it. The lowest $W_{\rm 80}$ values are observed to the southeast of the nucleus for H$_2$, while for [Ne\:{\sc iii}] they are seen to the northwest. Higher [Ne\:{\sc iii}] velocity dispersions are observed in a stripe perpendicular to the radio jet, centered approximately 2 arcsec west of the nucleus. This structure is interpreted as a biconical outflow by \citet{Dasyra15}. Higher $W_{\rm 80}$ values perpendicular to the jet orientation are also observed for H$_2$, but these are centered on the nucleus.  The H$_2$/PAH and [Ne\,\textsc{iii}]/[Ne\,\textsc{ii}] line ratios show very different behaviors: while the former displays the highest values within the inner $\sim$2 arcsec, the latter increases from the center outward along the jet direction, with lower values observed perpendicular to the jet. We also find higher temperatures for both the high- and low-excitation H$_2$ lines in regions close to the center and coincident with strong radio emission, similar to what was reported by \citet{Dasyra15} and attributed to shock excitation due to jet–cloud interactions. Evidence of shocked gas emission is also observed in the extended high-ionization gas, which shows high temperatures ($\sim20\,000$\,K), consistent with shock excitation produced by the interaction between the radio jet and the ISM \citep{Fonseca-Faria23}.

\noindent{\bf NGC\:1052} (Fig.~\ref{fig:maps_NGC1052}): NGC~1052 exhibits a radio jet with a total extent of 32~arcsec (2.8~kpc), oriented approximately along PA~$\approx 100^\circ$ , with the eastern radio knot located at $\sim 12$~arcsec from the nucleus and the western knot at $\sim 8$~arcsec \citep{Wrobel84}. Between these knots and the nucleus, fainter knots are observed at distances of $\sim 3-5$~arcsec on both sides of the nucleus, with a slightly different orientation (PA~$= 85^\circ$), which is the value we adopt as a reference in Table~\ref{tab:sample} \citep{Kadler04}. A more compact jet, on milliarcsecond scales, is oriented along PA~$= 70^\circ$, as revealed by very long-baseline interferometry observations \citep{Kadler04b}. 

A detailed analysis of the nuclear IR spectrum of NGC\,1052 using JWST NIRSpec IFU and MIRI/MRS data is presented in \citet{Goold26}, whereas spatially resolved emission-line maps derived from the MIRI/MRS observations are presented in \citet{Goold24}.  These authors provide flux and kinematic maps for the H$_2$ S(3), [Ar\,{\sc ii}]\,6.99\,$\mu$m, and [Ne\,{\sc iii}] emission lines, along with an analysis of the nuclear properties of different emission lines, which indicate the presence of ionized gas outflows tentatively associated with the direction of the compact jet \citep[see also][]{Cazzoli22}.

 The H$_2$ flux distribution in this galaxy is more elongated along the northeast–southwest direction, with lower intensity levels detected across the entire MRS FoV, similar to the behavior observed in the hot H$_2$ near-IR emission lines \citep{Dahmer-Hahn19}. A similar behavior is also observed in the PAH flux distribution. The [Fe\,{\sc ii}] emission is detected within the inner 2 arcsec radius and appears slightly more elongated along the same direction as the highest intensities of the H$_2$ and PAH emissions.
 The [Ne\,{\sc iii}] emission exhibits a centrally concentrated, nearly circular flux distribution. The H$_2$ $W_{\rm 80}$ map shows lower values to the north of the nucleus and higher values to the south of it, with overall lower values than those seen in ionized gas. The [Ne\,{\sc iii}] emission exhibits $W_{\rm 80}$ values of $\sim 1000$ km\,s$^{-1}$ at the nucleus and extending toward the west, following the orientation of the radio jet. In addition, a partial ring of lower $W_{\rm 80}$ values is observed at a radius of $\sim 2$ arcsec (also present in H$_2$, though less clearly) approximately co-spatial with a ring of reduced stellar population ages, as revealed by optical and near-IR IFU observations \citep{Dahmer-Hahn19b,Rogerio22}. The H$_2$/PAH and [Ne\,{\sc iii}]/[Ne\,{\sc ii}] ratio maps show similar distributions, with the lowest values at the nucleus and increasing toward the outer regions, primarily along the east-west direction, in a pattern similar to that seen in the [O\,{\sc iii}]$\lambda5007$/H$\beta$ map presented by \citet{Dahmer-Hahn19}, which traces the AGN ionization cone. The H$_2$ temperature maps show higher values in regions close to the nucleus and lower values at larger distances. However, the highest values of $T_{\rm H_2(3,1)}$ are observed slightly north of the nucleus, while those of $T_{\rm H_2(5,3)}$ are offset to the south.

\noindent{\bf M\:87} (Fig.~\ref{fig:maps_M87}): The galaxy M\:87 is well known for hosting the supermassive black hole M87*, whose shadow was captured for the first time by the Event Horizon Telescope, as a bright, asymmetric emission ring with an angular size of $d = 42 \pm 3~\mu$as \citep{EHT19}. From this compact region, a radio jet emerges, well collimated and bright, approximately oriented in the northwest direction (PA=290$^\circ$), exhibiting several emission knots from sub-parsec scales up to a few kiloparsecs \citep[e.g.,][]{Biretta91,Biretta95,Walker18}. On larger scales, of up to $\sim 40$ kpc, the jet exhibits sharper bends and irregular structures, with filamentary features apparent along its length, consistent with a FR I class \citep{Biretta83,Biretta91,Owen89}. 

\citet{Goold26} presented the nuclear spectrum of M\,87 based on JWST NIRSpec IFU and MIRI/MRS observations as part of a study of a sample of eight local low-luminosity AGN host galaxies. For M\,87, the authors report the detection of only 13 emission lines, corresponding to the poorest emission-line spectrum among all sources in the sample. To date, spatially resolved results based on JWST observations have not yet been published for this galaxy. The PAH\,11.3$\:\mu$m emission feature is not detected in this galaxy. The H$_2$ and [Fe\:{\sc ii}] emissions are mostly confined to the inner 2 arcsec, with the [Fe\:{\sc ii}] flux distribution being centrally concentrated and roughly round, while the H$_2$ emission shows an elongation perpendicular to the jet orientation, along with some knots aligned with the jet. In contrast, the [Ne\,{\sc iii}] emission is detected across the entire MRS FoV and appears more elongated toward the northwest, consistent with the orientation of the ionized gas traced by optical emission lines on larger (kiloparsec) scales  in Verry Large Telescope (VLT) Multi Unit Spectroscopic Explorer (MUSE) observations \citep{Boselli19,Osorno23}. The gas velocity dispersion shows values above $W_{\rm 80} \sim 1000\:$km\:s$^{-1}$, likely associated with outflowing gas as reported in previous studies \citep{Osorno23}. The [Ne\,{\sc iii}]/[Ne\,{\sc ii}] ratio exhibits its lowest values at the nucleus and in regions beyond 2 arcsec from it, while the highest values are observed to the west of the nucleus. The H$_2$ excitation temperatures were measured only in the very central region, and no clear spatial variations could be identified.

\end{appendix}
\end{document}